%% file: aaai_v3.tex
\documentclass[11pt]{article}

\usepackage[margin=1in]{geometry}
\usepackage{newtxtext,newtxmath}
\usepackage[hyphens]{url}
\usepackage{graphicx}
\usepackage{natbib}
\usepackage{bibunits}
\usepackage{caption}
\usepackage{hyperref}
\usepackage[hyphens]{url}  
\usepackage{graphicx} 
\usepackage{natbib}  
\usepackage{caption} 
\usepackage{algorithm}
\usepackage{algpseudocode}
\usepackage{newfloat}
\usepackage{listings}
\DeclareCaptionStyle{ruled}{labelfont=normalfont,labelsep=colon,strut=off} 
\floatstyle{ruled}
\newfloat{listing}{tb}{lst}{}
\floatname{listing}{Listing}

\usepackage{booktabs}
\usepackage{amsmath}
\usepackage{amsfonts}
\usepackage{array}

\title{Client-Side Probing of Deleted Ridge Statistics in Federated Unlearning}

\author{
Yijun Quan \qquad Giovanni Montana\\[0.5em]
University of Warwick\\
\texttt{yijun.quan@warwick.ac.uk}\\
\texttt{g.montana@warwick.ac.uk}
}
\date{}
\begin{document}

\maketitle
\begin{bibunit}[plainnat]

\begin{abstract}
Federated unlearning aims to remove a client's data from a shared model without retraining from scratch. Some efficient systems make deletion exact by storing compact, additive summaries of the training features and broadcasting an updated linear classifier after every accepted change. We show that these broadcasts can also reveal the hidden summaries. A malicious client can submit known changes, use the returned classifiers to identify the server state, and compare states immediately before and after an isolated deletion. This exposes the deleted sample, class, or client summary and can enable its reinsertion. We characterize exactly when the observations contain enough independent information, give a matching optimal construction for unrestricted probes, and derive a more realistic estimator based on additions formed from the attacker's own data. On MNIST and CIFAR-10, high-precision broadcasts permit exact label recovery for every tested sample deletion with both probe types. Lower-precision broadcasts sharply reduce fine-grained recovery, and insufficiently diverse responses prevent identification altogether. Unrestricted probes are readily detected by their size; most individual attacker-data additions resemble honest batches, although we do not claim that the complete sequence is inconspicuous. The results identify a concrete privacy and integrity risk, its algebraic cause, and practical limits involving broadcast precision, update verification, response rate, and concurrent activity.
\end{abstract}
\section{Introduction}
Federated learning keeps raw data distributed across clients, but a participant may later request removal of a sample, a class, or its entire contribution. Federated unlearning (FU) \cite{liu2022right} seeks to remove that influence without collecting the raw data centrally or retraining the shared model from scratch. A useful FU mechanism must therefore combine data locality and efficient deletion with fidelity to retraining on the retained data.

A common efficient design keeps a large pretrained feature extractor fixed and updates only a lightweight linear classifier. The classifier can be recomputed from two compact sums: products between training features, and products between features and labels. These sums are \emph{sufficient statistics}: they contain all information that this classifier needs from the retained training set. Because clients can add or subtract their contributions, the approach supports exact continual learning and unlearning without starting over \cite{quan2026exact}. We ask what an active client can learn when the server accepts such updates and returns the new classifier after each one.

Attacks that compare models before and after unlearning already show that deletion can expose data. Gao et al. formalize deletion inference and reconstruction across several learning tasks \cite{gao2022deletion}; Hu et al. recover internal features with model access and infer labels through prediction queries \cite{hu2024learn}; and Bertran et al. reconstruct samples from regularized linear models with fixed feature extractors \cite{bertran2024reconstruction}. Bertran et al. estimate a hidden training-data quantity using an independent public sample from the same distribution; our attack instead identifies the required server summary from its responses. Zhou et al. similarly invert models observed before and after federated unlearning from a server adversary \cite{11400570}. Those attacks use the model change caused by deletion, whereas our participating client first identifies the server state and then recovers the deleted aggregate.

Related work also studies malicious clients and deletion requests. Federated clients can reconstruct peers from consecutive training updates or improve reconstruction through malicious updates \cite{wilson2024federated,yue2025byzantine}. Sheng et al. infer and reconstruct unlearned data, then use those reconstructions either to impede forgetting or to degrade performance on the deleted data \cite{sheng2026retaliatory}. Huang et al. show that unlearning requests for samples never present in training can destroy model accuracy \cite{huang2025unlearn}. Cohen et al. prove that, for certain tasks, an exact unlearning mechanism can let an adversary controlling few points reconstruct almost the entire dataset through deletion requests \cite{cohen2026protecting}. Their result is a general information-theoretic warning. We give a constructive attack for systems that store the additive summaries introduced above and publish the complete classifier, together with an exact success condition and measured precision limits.

The attack exploits the information revealed across multiple valid interactions, not a single before-and-after pair. The client submits changes whose effect it knows, observes how the classifier moves, and repeats until the responses determine the server's regularized training summary. It performs this identification immediately before and after one isolated deletion. The difference reveals the deleted aggregate even when the regularization strength is unknown, because that unchanged term cancels. Resubmitting the recovered aggregate then reverses the deletion. We study both unrestricted matrix changes, which expose the algebraic limit of the interface, and additions assembled from the attacker's own examples.

Our contributions connect the threat model, theory, and measured attack. First, we prove a necessary-and-sufficient condition for exact state identification: the collected classifier changes must span the full feature space. We give a probe construction that meets the resulting minimum number of responses. Second, we derive the corresponding estimator for cumulative additions made from attacker data and show why owning enough feature-diverse examples is necessary but not sufficient. Third, we recover and replay isolated sample-, class-, and client-level deletions. Experiments with two image datasets, two feature extractors, and two broadcast precisions measure where the attack succeeds, where numerical error overwhelms small deletions, and which messages a simple size-based detector can flag.

\section{Background}
\label{sec:bg}
The studied federated continual-unlearning protocol~\cite{quan2026exact} exchanges two matrix summaries between each client and the server. Let $\phi(\cdot)$ denote a frozen feature extractor shared by all clients, let $d$ be its feature dimension, and let $c$ be the number of classes. Client $k$ maintains a local retained dataset $D_{k,t}$ at round $t$ with $D_{k,0}=\emptyset$. Between rounds, it may receive an add batch $D_{k,t}^{+}$ and a delete batch $D_{k,t}^{-} \subseteq D_{k,t-1}$. Applying $\phi(\cdot)$ gives $F_{k,t}^{\pm}\in\mathbb{R}^{n_{k,t}^{\pm}\times d}$ and one-hot label matrices $Y_{k,t}^{\pm}\in\{0,1\}^{n_{k,t}^{\pm}\times c}$. The client then computes the feature-product matrix $S_{k,t}^{\pm}\in\mathbb{R}^{d\times d}$ and feature--label matrix $G_{k,t}^{\pm}\in\mathbb{R}^{d\times c}$ as
\begin{equation}
S_{k,t}^{+} = (F_{k,t}^{+})^\top F_{k,t}^{+}, \qquad
G_{k,t}^{+} = (F_{k,t}^{+})^\top Y_{k,t}^{+},
\end{equation}
and
\begin{equation}
S_{k,t}^{-} = (F_{k,t}^{-})^\top F_{k,t}^{-}, \qquad
G_{k,t}^{-} = (F_{k,t}^{-})^\top Y_{k,t}^{-}.
\end{equation}
The client transmits $\big(S_{k,t}^{+}, G_{k,t}^{+}, S_{k,t}^{-}, G_{k,t}^{-}\big)$ to the server.

At the server, the global add and delete statistics are aggregated as
\begin{equation}
S_t^{+} = \sum_{k=1}^{K} S_{k,t}^{+}, \qquad
G_t^{+} = \sum_{k=1}^{K} G_{k,t}^{+},
\label{eqn:add}
\end{equation}
and
\begin{equation}
S_t^{-} = \sum_{k=1}^{K} S_{k,t}^{-}, \qquad
G_t^{-} = \sum_{k=1}^{K} G_{k,t}^{-}.
\label{eqn:sub}
\end{equation}
The server maintains a retained-statistics ledger,
\begin{equation}
S_t \leftarrow S_{t-1} + S_t^{+} - S_t^{-}, \qquad
G_t \leftarrow G_{t-1} + G_t^{+} - G_t^{-},
\end{equation}
and recovers the ridge-head parameters in closed form by
\begin{equation}
W_t = (S_t + \gamma I)^{-1}G_t,
\end{equation}
where $W_t\in\mathbb{R}^{d\times c}$ and $\gamma > 0$ is a ridge coefficient. A logical probe comprises all constituent summaries that create one chosen net perturbation and the single full-head broadcast after the server applies them. Client messages, constituent add/delete pairs, and server broadcasts are counted separately in the evaluation.

At each round, a client observes the current global weight $W_t$ and its own local summaries, but not other clients' data, features, gradients, or statistics. A returned head alone does not identify $(S_t,G_t)$ because the mapping $(S_t,G_t)\mapsto W_t=(S_t+\gamma I)^{-1}G_t$ is many-to-one. The attack below resolves this ambiguity by actively collecting responses to known summary changes.

\section{Threat Model}
Passive observation of $W_t$ does not identify $(S_t,G_t)$ because many ridge states produce the same head. We therefore consider an active malicious client that submits chosen sufficient-statistic summaries and observes the full head returned by the server after each logical probe. The client knows the frozen encoder and its own messages but not other clients' data, statistics, or the server's fixed $\gamma$.

The protocol guarantees the same result as retraining only when each deletion belongs to the requesting client's retained data. We test a server that does not verify record ownership, message origin, or duplicate submissions. The attack also needs a quiet interval: no other client may update the state while each sequence runs. Exactly one deletion of the stated size must occur between the two sequences, although its content remains unknown. Ownership checks can stop fabricated probes and unauthorized replay. They do not stop an attacker from learning through valid additions. Defending that channel requires fewer broadcasts, batched responses, partial classifiers, or added noise.

\subsection{Exact Identification from Moment Differences}
\label{sec:exact_identification}
Immediately before probing, define
\begin{equation}
A=S+\gamma I\succ0,\qquad H=A^{-1},\qquad W_0=HG.
\label{eqn:baseline_state}
\end{equation}
For probe response $j$, let $Q_j\in\mathbb{R}^{d\times c}$ be the known \emph{total} moment perturbation present when $W_j$ is observed, with no net Gram perturbation. If each probe is restored before the next, $Q_j=\Delta G_j$; if probes accumulate, $Q_j=\sum_{\ell=1}^{j}\Delta G_\ell$. In both cases,
\begin{equation}
R_j:=W_j-W_0=A^{-1}Q_j,\qquad AR_j=Q_j.
\label{eqn:response_difference}
\end{equation}
Stack the response differences and known perturbations as
\begin{equation}
R=[R_1,\ldots,R_m],\quad
Q=[Q_1,\ldots,Q_m],
\label{eqn:stacked_probes}
\end{equation}
so that $R,Q\in\mathbb{R}^{d\times mc}$ and
\begin{equation}
R=A^{-1}Q=HQ,\qquad AR=Q.
\label{eqn:two_recovery_systems}
\end{equation}

\paragraph{Realizing a moment-only probe.}
An arbitrary moment change can be expressed through the protocol's add/delete matrices while leaving the Gram matrix unchanged. For a feature vector $u$ and one-hot label $y$,
\begin{equation}
uu^\top-(-u)(-u)^\top=0,\qquad
uy^\top-(-u)y^\top=2uy^\top.
\label{eqn:algebraic_pair}
\end{equation}
Applying this construction to each class column realizes any desired $Q_j\in\mathbb{R}^{d\times c}$. Each submitted Gram matrix is positive semidefinite and each moment has one-hot-label form. The construction is algebraically valid, but the vector $-u$ need not be produced by the shared encoder and the deletion need not pass an ownership check. The data-derived variant below avoids fabricated feature vectors, although our evaluated sequence still assumes that duplicate submissions are accepted.

\paragraph{Theorem 1 (exact recovery from moment probes).}
Fix $\gamma>0$, which need not be known to the attacker, and consider any protocol whose hidden additive state satisfies $A=S+\gamma I\succ0$, $S\succeq0$, $G\in\mathbb{R}^{d\times c}$, and whose released head is $W=A^{-1}G$. If its interface accepts the moment-only changes above and releases the resulting full head, then, in exact arithmetic, the observations $(W_0,\{W_j,Q_j\}_{j=1}^{m})$ uniquely identify $(A,G)$ if and only if
\begin{equation}
\operatorname{rank}(Q)=d,
\label{eqn:rank_condition}
\end{equation}
equivalently $\operatorname{rank}(R)=d$. When this holds,
\begin{equation}
A=QR^\dagger,\qquad H=RQ^\dagger,\qquad G=AW_0,
\label{eqn:exact_recovery}
\end{equation}
where $\dagger$ denotes the Moore--Penrose pseudoinverse. Thus the regularized Gram matrix needed for deletion recovery is obtained directly, without inverting an estimated $\widehat H$.

For sufficiency, $A$ is invertible, so $\operatorname{rank}(R)=\operatorname{rank}(Q)$. At full row rank, right-multiplying $AR=Q$ by $R^\dagger$ gives $A=QR^\dagger$; similarly, $R=HQ$ gives $H=RQ^\dagger$, followed by $G=AW_0$. For necessity, suppose $\operatorname{rank}(R)<d$ and choose nonzero $v$ with $v^\top R=0$. For any $\varepsilon>0$, let
\begin{equation}
A'=A+\varepsilon vv^\top,\qquad
S'=S+\varepsilon vv^\top\succeq0,\qquad
G'=A'W_0.
\end{equation}
Then $A'=S'+\gamma I\succ0$, $A'\ne A$, and $A'R=AR=Q$. Hence $A'W_j=A'(W_0+R_j)=G'+Q_j$ for every probe, while $A'W_0=G'$. A distinct feasible ridge state therefore produces exactly the same observations. Necessity is over this algebraic state class; it does not require the alternative $G'$ to arise from a particular labelled dataset.

\paragraph{Scope beyond the studied protocol.}
RanPAC \cite{mcdonnell2023ranpac} independently maintains $\mathcal G=\sum hh^\top$ and $C=\sum hy^\top$ for frozen projected features and forms $W_o=(\mathcal G+\lambda I)^{-1}C$. This matches the algebraic state above. RanPAC is neither federated unlearning nor an attack target here; Theorem~1 reaches another deployment only if it also accepts known changes and releases every full head.

\paragraph{Corollary 1 (exact logical-response complexity).}
Since each $Q_j$ has at most $c$ independent columns, full row rank requires
\begin{equation}
m\ge \left\lceil\frac{d}{c}\right\rceil.
\label{eqn:probe_lower_bound}
\end{equation}
The bound is achievable when arbitrary algebraic moment probes are accepted: for $m=\lceil d/c\rceil$, partition the columns of $Q=[\tau I_d,0]\in\mathbb{R}^{d\times mc}$, with $\tau>0$, into $m$ blocks. Then $QQ^\top=\tau^2I_d$, so $Q$ has full row rank, $\sigma_{\min}(Q)=\tau$, and condition number one. Therefore the noiseless chosen-summary query complexity is exactly $\lceil d/c\rceil$ logical full-head responses per unknown state. This statement does not exploit additional dataset-realizability structure.

With 512-dimensional features and ten classes, identifying one unknown state needs 52 probe responses plus its baseline. A first attack identifies the states on both sides of a deletion. It therefore uses 104 probe responses and 109 server responses in total. The other five responses are the initial baseline, two cancellations, the deletion, and replay. Once a pre-deletion state has been identified and restored, a later event needs 55 responses.

These counts treat one complete matrix block as a logical probe and assume one broadcast after that block. A server that broadcasts after every rank-one part would return many more responses. We therefore report logical probes, client messages, and server broadcasts separately.

The attacker first records a baseline and identifies the pre-deletion state, then cancels its cumulative probe. After observing one isolated deletion, it treats the resulting head as the post-deletion baseline, identifies that state, and cancels the second probe sequence. The difference between the two identified states yields the deleted aggregate block, which the attacker can then replay.

With an observed response stack $\widetilde R=R+E$ and full-row-rank $\widetilde R$, direct recovery
\begin{equation}
\widehat A=Q\widetilde R^\dagger
\label{eqn:finite_precision_bound}
\end{equation}
satisfies
\begin{equation}
\widehat A-A=-AE\widetilde R^\dagger,\qquad
\|\widehat A-A\|_F
\le\frac{\|A\|_2\|E\|_F}{\sigma_{\min}(\widetilde R)}.
\label{eqn:direct_a_error}
\end{equation}
Here $E$ includes error in each probe response and in the shared baseline subtracted from every response difference. The complementary estimator $\widehat H=\widetilde RQ^\dagger$ obeys
$\|\widehat H-H\|_F\le\|E\|_F/\sigma_{\min}(Q)$ when $Q$ has full row rank. The implementation therefore uses SVD- or QR-based solves and reports the numerical rank and conditioning of both $Q$ and $\widetilde R$, together with direct-$A$ and direct-$H$ agreement.

If the baseline is observed as $\widetilde W_0=W_0+N_0$, then $\widehat G=\widehat A\widetilde W_0$ has error
\begin{equation}
\widehat G-G=(\widehat A-A)W_0+\widehat A N_0.
\label{eqn:moment_error_propagation}
\end{equation}
Thus moment recovery depends on both state-identification error and baseline precision. We solve for $A$ directly, symmetrize the estimate, and reject it unless it is positive definite. An independently estimated $H=A^{-1}$ is also symmetrized and checked for positive definiteness and agreement with $\widehat A$; it is a diagnostic and is never inverted to obtain $A$. The supplementary material reports the numerical residuals. The recovered moment is $\widehat G=\widehat A W_0$.

In the cumulative implementation, each returned classifier is paired with the total perturbation present at that time. The attacker cancels the accumulated perturbation after the last response. Numerical rank counts singular values above $\varepsilon_{\mathrm{rank}}\sigma_{\max}$. The full pseudocode, independent inverse estimate, definiteness checks, and solve diagnostics appear in the supplement.

\subsection{Attacker-Data Additions}
The unrestricted construction may use feature vectors that the shared encoder cannot produce. We therefore also study additions formed from the attacker's own examples. The additions accumulate until the server returns classifier $W_j$. Let $(P_j,Q_j)$ denote their known total feature-product and feature--label summaries. Subtracting the baseline equation gives
\begin{equation}
A(W_j-W_0)=Q_j-P_jW_j.
\label{eqn:genuine-probe}
\end{equation}
Stack the observed changes as $X=[W_1-W_0,\ldots,W_m-W_0]$ and the known right-hand sides as $Z=[Q_1-P_1W_1,\ldots,Q_m-P_mW_m]$. The state is identifiable exactly when $X$ has rank $d$, in which case $A=ZX^\dagger$.

If probe $j$ contains features $F_j$ and labels $Y_j$, then $Z_j=F_j^\top(Y_j-F_jW_j)$. Let $F_{\mathrm{attacker}}$ stack all distinct attacker features used by the sequence. Then $\operatorname{rank}(X)=\operatorname{rank}(Z)\le\operatorname{rank}(F_{\mathrm{attacker}})$. The attacker therefore needs at least $d$ examples and full feature rank. These conditions are not sufficient because the prediction residuals and batching sequence also affect $Z$. Thus 52 responses is only a dimensional lower bound for attacker-data additions.

\subsection{Recovering and Replaying an Isolated Deletion}
Let $(A_{\mathrm{pre}},W_{\mathrm{pre}})$ and $(A_{\mathrm{post}},W_{\mathrm{post}})$ be two identified states bracketing one deletion, with the same fixed $\gamma$. The deleted contribution is exactly
\begin{equation}
S_{\mathrm{del}}=A_{\mathrm{pre}}-A_{\mathrm{post}},\qquad
G_{\mathrm{del}}=A_{\mathrm{pre}}W_{\mathrm{pre}}
-A_{\mathrm{post}}W_{\mathrm{post}}.
\label{eqn:deleted_update}
\end{equation}
Indeed, deletion gives $A_{\mathrm{post}}=A_{\mathrm{pre}}-S_{\mathrm{del}}$ because the same $\gamma$ appears in both states, while $G_{\mathrm{pre}}=A_{\mathrm{pre}}W_{\mathrm{pre}}$ and $G_{\mathrm{post}}=A_{\mathrm{post}}W_{\mathrm{post}}$. Adding the recovered pair to the post-deletion ledger therefore returns both sufficient statistics and the head exactly in exact arithmetic. If several honest events occur between the two identified states, the equations recover only their signed aggregate net change and cannot attribute it to a particular event or client. If $\gamma$ changes between states, the Gram difference is additionally shifted by $(\gamma_{\mathrm{pre}}-\gamma_{\mathrm{post}})I$.
After the cumulative probes are cancelled, the implementation records the numerical consistency diagnostic
\begin{equation}
r_W=\frac{\|W_{\mathrm{final}}-W_0\|_F}
{\max(\|W_0\|_F,\varepsilon_{\mathrm{den}})},
\label{eqn:consistency_check}
\end{equation}
with $\varepsilon_{\mathrm{den}}=10^{-15}$. Rank and the two positive-definiteness tests determine identification success; $r_W$ is reported but is not a success criterion. Equality of heads does not certify equality of hidden states, so this diagnostic does not prove that no honest client updated. The supplementary material gives simulator-only hidden-ledger checks.

To connect state-estimation error to replay integrity, write
$\widehat A_{\mathrm{pre}}=A_{\mathrm{pre}}+E_{\mathrm{pre}}$ and
$\widehat A_{\mathrm{post}}=A_{\mathrm{post}}+E_{\mathrm{post}}$.
The replayed head then satisfies the exact identity
\begin{equation}
\begin{aligned}
W_{\mathrm{replay}}-W_{\mathrm{pre}}
&=\bigl(A_{\mathrm{pre}}+E_{\mathrm{pre}}-E_{\mathrm{post}}\bigr)^{-1}\\
&\quad E_{\mathrm{post}}(W_{\mathrm{pre}}-W_{\mathrm{post}}),
\end{aligned}
\label{eqn:replay_error_identity}
\end{equation}
whenever the leading matrix is invertible. Consequently,
\begin{equation}
\begin{aligned}
\|W_{\mathrm{replay}}-W_{\mathrm{pre}}\|_F
&\le \|\bigl(A_{\mathrm{pre}}+E_{\mathrm{pre}}-E_{\mathrm{post}}\bigr)^{-1}\|_2 \\
&\quad\times\|E_{\mathrm{post}}\|_2
\|W_{\mathrm{pre}}-W_{\mathrm{post}}\|_F.
\end{aligned}
\label{eqn:replay_error_bound}
\end{equation}
The pre-state error therefore enters only through the inverse, whereas the post-state error also controls the numerator. In particular, an exact post-state estimate can restore the head even when the replayed ledger remains inaccurate. If the observed baselines contain errors $N_{\mathrm{pre}}$ and $N_{\mathrm{post}}$, the numerator additionally contains $\widehat A_{\mathrm{pre}}N_{\mathrm{pre}}-\widehat A_{\mathrm{post}}N_{\mathrm{post}}$.

Matching the earlier classifier does not prove that the hidden summaries were restored: distinct server states can return the same classifier. The supplement proves this fact and gives the evaluator-only checks used for the feature-product summary, feature--label summary, and returned classifier. It also reports the smallest eigenvalue of the recovered deleted Gram block, because finite-precision error can violate the positive-semidefinite constraint expected by a validating server.

\section{Probing Attack Evaluation}

We test three consequences of the attack. First, can it recover the label and feature of one deleted sample? Second, can it recover the combined summaries removed by a class or client deletion? Third, can replaying those summaries reverse the deletion?

Our primary experiments map images through a frozen ImageNet-1K ResNet-18 \cite{He_2016_CVPR,5206848} to 512-dimensional features. A robustness check instead uses the frozen 768-dimensional DINOv2 ViT-B/14 encoder \cite{oquab2024dinov2}. We use MNIST \cite{6296535} and CIFAR-10 \cite{krizhevsky2009learning}, fixed $\gamma=10^{-3}$, 50 clients, and a Dirichlet split with concentration $\alpha=0.05$, which produces highly uneven class proportions across clients. Every evaluated split contains at least ten samples per client. Server ledgers and solves remain float64; the primary setting returns float64 heads and the precision ablation rounds every returned head and baseline to float32.

The simulated server updates its retained summaries and returns the classifier after solving the ridge system. An attack first identifies the pre-deletion state and cancels its probes. The server then applies one honest deletion. The returned classifier becomes the post-deletion baseline. The attack identifies that state, cancels again, recovers the difference, and replays it from the actual post-cancellation state.

Recovery uses only returned classifiers and known attacker messages. The evaluator reads the hidden summaries only to apply the chosen deletion and score recovery, cancellation, and replay. Failures remain in all success denominators. Error averages include successful runs only.

We measure recovery of the deleted moment and Gram matrices with relative Frobenius error:
\[
\mathrm{RelErr}(\Delta G)
=
\frac{\|\widehat{\Delta G} - \Delta G\|_F}
     {\|\Delta G\|_F},
\]
\[
\mathrm{RelErr}(\Delta S)
=
\frac{\|\widehat{\Delta S} - \Delta S\|_F}
     {\|\Delta S\|_F}.
\]

\paragraph{Rank threshold.}
Across five random probe sequences per dataset (Table~\ref{tab:rank-threshold}), $m=51$ always gives rank 510 and fails. At $m=52$, every stack reaches rank 512 and succeeds. The designed 52-response probe also reaches rank 512. Its float64 state error is $1.51\times10^{-12}$ on MNIST and $6.55\times10^{-13}$ on CIFAR-10.

\begin{table*}[t]
\centering
\caption{Float64 identification of one hidden server state. Numerical rank uses relative tolerance $10^{-10}$. Random probes use independent Gaussian moment increments scaled by $10^4$ and report mean $\pm$ standard deviation over five sequences. The designed probe uses $Q=[10^4I,0]$. Every rank-510 run fails and every rank-512 run succeeds.}
\label{tab:rank-threshold}
\footnotesize
\setlength{\tabcolsep}{4pt}
\begin{tabular}{@{}lllccccc@{}}
\toprule
Data & Probe & $m$ & Success & $\operatorname{rank}(Q)$ & $\kappa(Q)$ & $\kappa(R)$ & $\mathrm{RelErr}(A)$ \\
\midrule
MNIST & Random & 51 & $0/5$ & 510 & $\infty$ & $\infty$ & -- \\
 & Random & 52 & $5/5$ & 512 & $(5.35\!\pm\!.38)\times10^3$ & $(6.79\!\pm\!1.46)\times10^7$ & $(8.84\!\pm\!11.65)\times10^{-11}$ \\
 & Random & 64 & $5/5$ & 512 & $(5.08\!\pm\!.33)\times10^2$ & $(2.22\!\pm\!.39)\times10^7$ & $(2.44\!\pm\!1.63)\times10^{-11}$ \\
 & Designed & 52 & $1/1$ & 512 & 1 & $1.06\times10^6$ & $1.51\times10^{-12}$ \\
\addlinespace
CIFAR-10 & Random & 51 & $0/5$ & 510 & $\infty$ & $\infty$ & -- \\
 & Random & 52 & $5/5$ & 512 & $(5.35\!\pm\!.38)\times10^3$ & $(4.05\!\pm\!.71)\times10^6$ & $(8.13\!\pm\!4.32)\times10^{-12}$ \\
 & Random & 64 & $5/5$ & 512 & $(5.08\!\pm\!.33)\times10^2$ & $(1.23\!\pm\!.11)\times10^6$ & $(1.66\!\pm\!1.47)\times10^{-12}$ \\
 & Designed & 52 & $1/1$ & 512 & 1 & $4.15\times10^4$ & $6.55\times10^{-13}$ \\
\bottomrule
\end{tabular}
\end{table*}

\paragraph{Probe scale.}
Every tested MNIST scale succeeds (see the supplementary material), but the designed probe is conspicuous. At $\tau=10^4$, its message norm is $2.98\times10^5$ times the honest-update 99th percentile. The largest classifier change is $8.15\times10^4$ times its honest threshold. Clipping each payload to the honest threshold still produces a 149-fold classifier change. Deleted-block errors remain $3.68\times10^{-6}$ for $S$ and $1.76\times10^{-6}$ for $G$. The supplement gives the full sweep.

\begin{table*}[t]
\centering
\caption{Float64 deletion recovery with the designed probe ($m=52$ responses per state, excluding the baseline). Errors are mean $\pm$ standard deviation over successful attacks only. Client results include all 50 deletions in each of 20 valid client partitions; failures remain in every success denominator.}
\label{tab:results}
\small
\begin{tabular}{@{}llcccc@{}}
\toprule
Deletion & Dataset & Successful & Labels & $\mathrm{RelErr}(\Delta G)$ & $\mathrm{RelErr}(\Delta S)$ \\
\midrule
Sample & MNIST & $100/100$ & $100/100$ & $(4.14 \pm 2.67)\times 10^{-8}$ & $(1.16 \pm 0.83)\times 10^{-7}$ \\
       & CIFAR-10 & $100/100$ & $100/100$ & $(1.06 \pm 0.48)\times 10^{-8}$ & $(2.69 \pm 1.52)\times 10^{-8}$ \\
\addlinespace
Class  & MNIST & $10/10$ & -- & $(5.97 \pm 4.28)\times 10^{-12}$ & $(1.67 \pm 1.77)\times 10^{-11}$ \\
       & CIFAR-10 & $10/10$ & -- & $(1.55 \pm 1.12)\times 10^{-12}$ & $(4.24 \pm 3.62)\times 10^{-12}$ \\
\addlinespace
Client & MNIST & $1000/1000$ & -- & $(3.87 \pm 10.99)\times 10^{-10}$ & $(9.14 \pm 25.78)\times 10^{-10}$ \\
       & CIFAR-10 & $1000/1000$ & -- & $(7.69 \pm 21.87)\times 10^{-11}$ & $(1.80 \pm 4.99)\times 10^{-10}$ \\
\bottomrule
\end{tabular}
\end{table*}

Table~\ref{tab:results} reports the primary designed-probe results. All 2,220 float64 attacks identify both states. Sample labels are always recovered, while aggregate errors remain between $10^{-12}$ and $10^{-7}$ depending on deletion size.

Deletion recovery subtracts two estimates of the full state. Their absolute errors can be large relative to a small deleted block, so subtraction amplifies relative error most strongly for sample deletion.

For the deleted Gram block, let $E_{\mathrm{pre}}=\widehat A_{\mathrm{pre}}-A_{\mathrm{pre}}$ and $E_{\mathrm{post}}=\widehat A_{\mathrm{post}}-A_{\mathrm{post}}$. State differencing gives $\widehat{\Delta S}-\Delta S=E_{\mathrm{pre}}-E_{\mathrm{post}}$. Therefore
\begin{equation}
\mathrm{RelErr}(\Delta S)\le
B_S:=\frac{\|E_{\mathrm{pre}}\|_F+\|E_{\mathrm{post}}\|_F}{\|\Delta S\|_F}.
\label{eqn:gram-amplification-bound}
\end{equation}

\begin{figure}[t]
\centering
\includegraphics[width=.7\columnwidth]{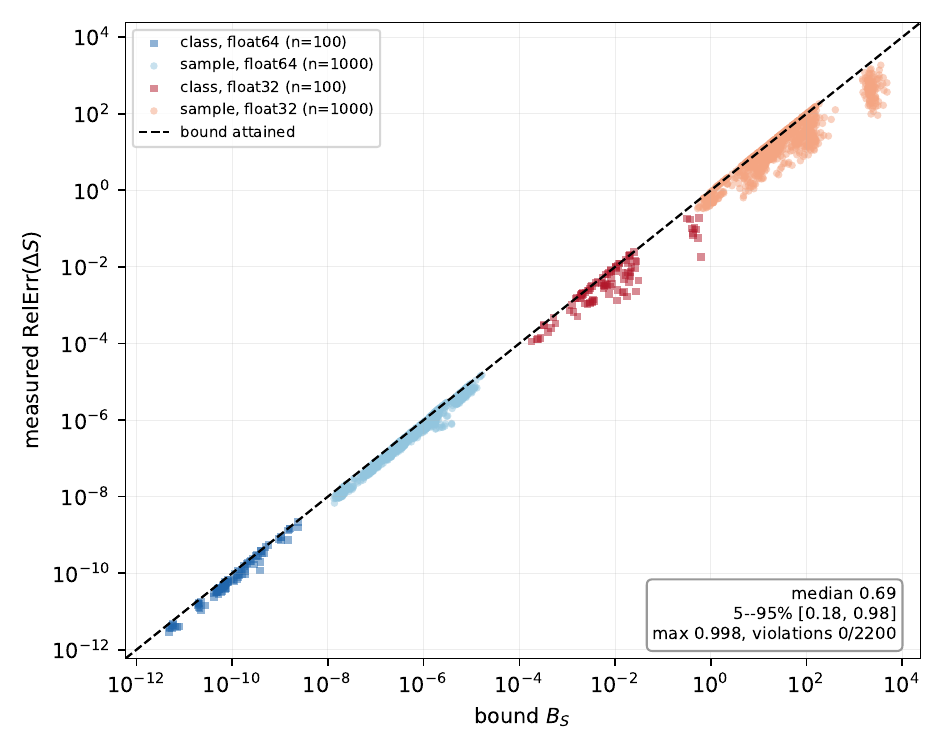}
\caption{Measured deleted-Gram error versus the bound in Equation~\eqref{eqn:gram-amplification-bound}. The 2,200 attacker-data attacks use 104 responses per state and cover both datasets, both precisions, and sample and class deletions. Every run shown passed both state-identification checks; this does not imply accurate deletion recovery. No point crosses the dashed bound. Measured-to-bound ratios have pooled median 0.69, 5--95\% range 0.18--0.98, maximum 0.998, and cell-median range 0.43--0.79.}
\label{fig:amplification-bound}
\end{figure}

Figure~\ref{fig:amplification-bound} checks this explanation run by run. The points follow the equality line across both precisions and deletion sizes. Precision changes the state error, while deletion size determines how strongly differencing amplifies it.

\begin{table*}[t]
\centering
\caption{Deletion recovery from $m=104$ cumulative attacker-data addition responses per state. At relative rank tolerance $10^{-10}$, every stack has full observed rank and both state estimates pass the positive-definiteness checks. ``$G/S$'' gives mean deleted-moment and deleted-Gram relative errors. Float32 changes only returned heads and baselines.}
\label{tab:genuine-results}
\small
\setlength{\tabcolsep}{5pt}
\begin{tabular}{@{}llcccc@{}}
\toprule
Data & Heads & Sample labels & Sample $G/S$ & Class $G/S$ & Client $G/S$ \\
\midrule
MNIST & float64 & $500/500$ & $1.55\times 10^{-6}/1.62\times 10^{-6}$ & $2.87\times 10^{-10}/3.73\times 10^{-10}$ & $1.08\times 10^{-8}/8.76\times 10^{-9}$ \\
CIFAR-10 & float64 & $500/500$ & $9.43\times 10^{-8}/1.24\times 10^{-7}$ & $2.09\times 10^{-11}/2.97\times 10^{-11}$ & $1.36\times 10^{-9}/1.85\times 10^{-9}$ \\
MNIST & float32 & $49/500$ & $68.4/143$ & $1.32\times 10^{-2}/2.74\times 10^{-2}$ & $0.393/0.661$ \\
CIFAR-10 & float32 & $183/500$ & $5.27/10.1$ & $1.16\times 10^{-3}/2.42\times 10^{-3}$ & $0.0842/0.161$ \\
\bottomrule
\end{tabular}
\end{table*}

\paragraph{Attacker data and cost.}
We instantiate the attacker-data additions using the attacker’s own examples. Across five splits, 25--33 of 50 MNIST clients and 25--28 CIFAR-10 clients pass the feature-rank gate. The selected attackers contain 867--4,309 and 779--5,342 distinct examples, respectively. We divide each set without overlap into 104 batches of 8--42 MNIST or 7--52 CIFAR-10 examples. The same set is reused after cancellation to identify the post-deletion state.

These examples already appear in the initial server ledger. The experiment therefore tests duplicate submissions of attacker-owned, encoder-produced data rather than held-out records. The interface does not reject duplicates. A complete first attack sends 208 additions, two cancellations, one deletion, and replay: 212 client messages and 213 server responses including the baseline. The supplement gives the attacker-selection rule.

\paragraph{Observed rank for attacker's data.}
With 52 additions, every pre-state reaches rank 512, but a class deletion leaves most post-state stacks deficient. Only $14/50$ MNIST and $16/50$ CIFAR-10 attacks succeed, and post-state ranks range from 468 to 512 (see supplement). At 104 additions, every tested stack passes at tolerance $10^{-10}$. Median $\kappa(X)$ is $4.06\times10^5$ on MNIST and $1.43\times10^5$ on CIFAR-10. Response count alone is therefore insufficient. Full attacker feature rank is necessary but not sufficient, and the bound does not require label diversity.

We also vary the float32 rank tolerance from $10^{-6}$ to $10^{-10}$ on ten pre- or post-class-deletion stacks per dataset. At $m=104$, all twenty stacks remain full rank from $10^{-7}$ through $10^{-10}$. At $10^{-6}$, all ten CIFAR-10 stacks but only six MNIST stacks remain full rank. Some MNIST directions are therefore marginal. More importantly, full numerical rank only means that the estimator returns; it does not imply accurate recovery.

A DINOv2 check reaches the same conclusion with 768-dimensional transformer features. None of the five stacks per dataset is full rank at the 77-response dimensional lower bound, whereas all are full rank at 104 responses. The supplement reports ranks, conditioning, and errors.

\paragraph{Individual-message detectability.}
We calibrate batch-matched 99th-percentile norm thresholds on 5,000 honest additions per dataset. Testing on 10,000 separate additions gives false-positive rates of 1.38\% on MNIST and 1.34\% on CIFAR-10. Median attacker-message ratios are 0.818 and 0.845 relative to these thresholds. None of 520 MNIST additions and 13 of 520 CIFAR-10 additions exceed them. This test excludes both cancellations and replay, so it does not show that the complete sequence is inconspicuous.

\paragraph{Broadcast precision.}
With float32 responses, attacker-data sample-label recovery falls to 9.8\% on MNIST and 36.6\% on CIFAR-10, versus 10\% chance. Seed-level rates are 7--14\% and 12--100\%, so the pooled CIFAR-10 rate hides strong state dependence. Larger class blocks remain more accurate. Designed probes are also precision-limited because their observed response stacks have condition numbers $1.06\times10^6$ and $4.15\times10^4$, despite a perfectly conditioned input probe. Rounding the final honest classifier to float32 changes none of 10,000 test predictions; accuracy remains 97.15\% and 85.94\%.

Figure~\ref{fig:amplification-bound} links the precision and deletion-size results through one mechanism. Float64 and float32 occupy different ranges of the same bound: response rounding increases state error, and differencing amplifies it most when the deleted block is small. The figure is a post-hoc bound check, not a measured prediction for untested numerical formats. Because the corresponding moment bound must also include baseline-rounding terms, we do not infer its tightness from the Gram result.

Finite precision also makes some recovered sample and client Gram blocks slightly non-positive-semidefinite. A server that validates this constraint would reject those raw replays. Clipping negative eigenvalues to zero makes every tested block valid. In float64, the resulting replay-head error remains below $6\times10^{-9}$. The supplement reports the full diagnostic table.

\subsection{Sample-level probing}

Sample-level experiment studies the most fine-grained privacy setting, in which one deleted training point is probed at a time. For a deleted sample $(f_i, y_i)$, the removed contribution to the sufficient statistics is
\begin{equation}
\Delta S_i = f_i f_i^\top, \qquad \Delta G_i = f_i y_i^\top.
\end{equation}
For a one-hot label $y_i=e_k$, the exact deleted moment is $\Delta G_i=f_i e_k^\top$: its unique nonzero column identifies $k$ and equals $f_i$. Numerically, small errors can make every column nonzero, so the implementation uses
\begin{equation}
\widehat k=\arg\max_{r\in\{1,\ldots,c\}}
\|\widehat{\Delta G}_{i,:,r}\|_2,\qquad
\widehat f_i=\widehat{\Delta G}_{i,:,\widehat k},
\label{eqn:sample_extraction}
\end{equation}
and reports $\|\widehat f_i-f_i\|_2/\|f_i\|_2$. If only $\Delta S_i=f_i f_i^\top$ were known, a nonzero $f_i$ would be identifiable up to a global sign, not an arbitrary orthogonal transformation.
Writing $\widehat{\Delta G}_i=f_i e_k^\top+E$, the maximum-column rule is guaranteed to return $k$ whenever
\begin{equation}
\|E_{:k}\|_2+\max_{r\ne k}\|E_{:r}\|_2<\|f_i\|_2;
\end{equation}
the simpler condition $\|E\|_F<\|f_i\|_2/2$ is sufficient.
For every correctly recovered label, the maximum-column rule implies
\begin{equation}
\frac{\|\widehat f_i-f_i\|_2}{\|f_i\|_2}
\le \operatorname{RelErr}(\Delta G_i).
\label{eqn:feature-error-check}
\end{equation}
Among correctly labelled float64 samples, mean relative feature error is $4.32\times10^{-7}$ on MNIST and $2.75\times10^{-8}$ on CIFAR-10. The supplementary material gives the complete feature-error summary. It also provides selected decoder outputs as a qualitative illustration. Each decoder was trained only on the selected attacker's local image--feature pairs. This test exposes the threat of possible content interpretation from the recovered features.

\subsection{Class-level probing}
Class-level experiment evaluates whether the probing attack can recover the aggregate statistics of a deleted class. Here, the deleted object is no longer a single point but the entire class-wise contribution to the retained-set ledger. If $\mathcal{D}_c$ denotes the set of all training samples from class $c$, then the attack seeks to reconstruct
\begin{equation}
\Delta S_c = \sum_{i \in \mathcal{D}_c} f_i f_i^\top, \qquad
\Delta G_c = \sum_{i \in \mathcal{D}_c} f_i y_i^\top. 
\end{equation}
This setting is especially important because it tests whether the probe can capture large, structured deletions rather than only isolated points.

Table~\ref{tab:results} reports small relative errors for the recovered class-level aggregate blocks.
Because all deleted labels equal $e_c$, the class moment has the form $\Delta G_c=(\sum_{i\in\mathcal D_c}f_i)e_c^\top$. When this feature sum is nonzero and distinguishable from numerical error, its column identifies the deleted class. The aggregate does not identify the individual examples in general.
Two branches are advanced sequentially: an honest branch accumulates ten class deletions, while the attacked branch replays each recovered class block immediately after its deletion. At the final step, their accuracies are 9.82\% versus 97.15\% on MNIST and 8.94\% versus 85.94\% on CIFAR-10. The attacked branch's final relative head errors from the original head are $2.42\times10^{-12}$ and $6.65\times10^{-13}$; full trajectories appear in the supplement. This measures reintroduction of aggregate blocks, not recovery of individual examples.

\subsection{Client-level probing}
Client-level deletion targets the aggregate contribution of dataset $\mathcal{D}_k$:
\begin{equation}
\Delta S_k = \sum_{i \in \mathcal{D}_k} f_i f_i^\top, \qquad
\Delta G_k = \sum_{i \in \mathcal{D}_k} f_i y_i^\top. 
\end{equation}
Across 20 independently sampled 50-client partitions per benchmark, Table~\ref{tab:results} reports all 1,000 deleted-client blocks. The honest branch accumulates deletions, whereas the attacked branch immediately replays every recovered block. Its mean final relative head errors are $(6.29\pm0.65)\times10^{-12}$ on MNIST and $(1.42\pm0.11)\times10^{-12}$ on CIFAR-10; the supplement gives the complete trajectories.

For a mixed-label client deletion, column $r$ of $\Delta G_k$ is the sum of deleted features with label $r$, while $\Delta S_k$ is their aggregate second moment. These aggregates reveal classwise feature sums and one overall second moment, but they do not identify individual examples in general.

\section{Discussion and Conclusions}
This work shows how repeated classifier releases can expose the compact training summaries used by exact ridge-based unlearning. We prove the precise condition for identifying one server state and match it with an optimal unrestricted probe. We then derive an estimator that uses additions made from attacker data. Identifying states immediately before and after a deletion reveals the deleted aggregate and enables replay.

The experiments show both success and clear limits. Every high-precision designed attack succeeds. With 104 attacker-data responses per state, every tested high-precision sample label is also recovered. The theoretical minimum response count does not guarantee success for attacker-data additions: some ResNet-18 and DINOv2 response stacks remain rank deficient. Lower precision sharply reduces sample recovery but leaves larger class aggregates more accurate. A simple norm detector catches the unrestricted probe but misses most individual attacker-data additions. We do not claim whether the complete sequence is stealthy.

Deletion size also changes the security consequence. Recovering one sample's feature and label is a confidentiality failure. A class or client aggregate need not reveal individual records, but replay can reverse the requested deletion. Float32 sharply reduces sample recovery while leaving class aggregates much more accurate, so it is not a uniform safeguard.

The probe types expose complementary weaknesses in simple defenses. A norm threshold detects the unrestricted probe, but with float32 responses it still recovers 92 of 100 MNIST labels and all 100 CIFAR-10 labels. Attacker-data additions usually pass that test and recover every label from float64 responses. 

The presented attack requires the complete classifier after each update and a quiet interval. It assumes fixed regularization and one isolated deletion of known size. The evaluated additions resubmit ledger records, so duplicate checking would block them. Finally, our identifiability conditions demand full feature rank.

These constraints point to concrete directions for securing ridge-based federated unlearning systems. Servers should verify ownership, reject duplicates, and authenticate replay. Valid new records may still reveal the state through repeated responses. Servers can respond less often, batch updates, release only part of the classifier, or add noise. Lower precision helps selectively. Future work should test held-out data under concurrency.

\putbib[ref] 
\end{bibunit}
\clearpage
\appendix
\input{aaai_v3_supplementary}
\bibliographystyle{plainnat}
\bibliography{ref}
\end{document}

%% file: aaai_v3_supplementary.tex


\onecolumn
\begin{center}
{\LARGE\bfseries Supplementary Material: Recovering and Replaying Deleted Ridge Statistics in Federated Unlearning\par}
\vspace{1em}
\end{center}
\vspace{1em}

\section{Exact Identification from Moment Probes}
\subsection{Notation and Observation Model}
Let the server state at one stable probing interval be
\begin{equation}
A=S+\gamma I\succ0,\qquad H=A^{-1}.
\label{eq:supp-baseline}
\end{equation}
Here $S\succeq0$, $G\in\mathbb{R}^{d\times c}$, and $W_0=A^{-1}G=HG$.
The attacker changes only the moment statistic. When response $W_j$ is observed, let $Q_j\in\mathbb{R}^{d\times c}$ be the known total moment perturbation currently present. Thus
\begin{equation}
R_j:=W_j-W_0=A^{-1}Q_j,\qquad AR_j=Q_j.
\label{eq:supp-response}
\end{equation}
For independently restored probes, $Q_j=\Delta G_j$. For cumulative probes,
\begin{equation}
Q_j=\sum_{\ell=1}^{j}\Delta G_\ell.
\label{eq:supp-cumulative}
\end{equation}
Define the horizontal stacks
\begin{equation}
\begin{aligned}
R&=[R_1,\ldots,R_m],\\
Q&=[Q_1,\ldots,Q_m],
\end{aligned}
\label{eq:supp-stacks}
\end{equation}
Here $R,Q\in\mathbb{R}^{d\times mc}$ and
\begin{equation}
R=A^{-1}Q=HQ,\qquad AR=Q.
\label{eq:supp-two-systems}
\end{equation}
\subsection{Exact Recovery Theorem}
\paragraph{Theorem 1 (exact recovery from moment probes).}
Fix any $\gamma>0$, which need not be known to the attacker, and consider
\[
\mathcal C_\gamma=\{(A,G):A=S+\gamma I,\ S\succeq0,\ G\in\mathbb R^{d\times c}\}.
\]
In exact arithmetic, the observations $(W_0,\{W_j,Q_j\}_{j=1}^{m})$ uniquely identify $(A,G)$ over $\mathcal C_\gamma$ if and only if
\begin{equation}
\operatorname{rank}(Q)=d.
\label{eq:supp-rank}
\end{equation}
Because $A$ is invertible, this is equivalent to $\operatorname{rank}(R)=d$. When the condition holds,
\begin{equation}
A=QR^\dagger,\qquad H=RQ^\dagger,\qquad G=AW_0.
\label{eq:supp-estimator}
\end{equation}
Here $\dagger$ denotes the Moore--Penrose pseudoinverse. When $\gamma$ is unknown, these observations identify $A=S+\gamma I$ rather than $S$ separately: every $0<\gamma'\le\lambda_{\min}(A)$ gives the feasible decomposition $S'=A-\gamma'I\succeq0$. A fixed $\gamma$ cancels when pre- and post-deletion values of $A$ are differenced.

\paragraph{Proof of sufficiency.}
Since $R=A^{-1}Q$, the stacks have the same rank. At full row rank, right-multiplying $AR=Q$ by $R^\dagger$ gives $A=QR^\dagger$; right-multiplying $R=HQ$ by $Q^\dagger$ gives $H=RQ^\dagger$; and then $G=AW_0$. Thus the state is uniquely determined without inverting an estimated $H$.

\paragraph{Proof of necessity.}
Suppose $\operatorname{rank}(R)<d$. Choose nonzero $v\in\mathbb R^d$ with $v^\top R=0$. For any $\varepsilon>0$, define
\[
A'=A+\varepsilon vv^\top,\qquad
S'=S+\varepsilon vv^\top\succeq0,\qquad
G'=A'W_0.
\]
Then $A'=S'+\gamma I\succ0$, $A'\ne A$, and $A'R=AR+\varepsilon vv^\top R=Q$. Therefore, for every $j$,
\[
A'W_j=A'(W_0+R_j)=G'+Q_j,
\]
while $A'W_0=G'$. The distinct valid ridge state $(A',G')$ produces the same baseline and all probe responses, so unique identification is impossible. This necessity statement is over the stated algebraic class; $G'$ need not be jointly realizable by a particular labelled dataset.
\subsection{Exact Logical-Response Complexity}
If a logical response follows a total probe of rank at most $r_{\max}$, then
\begin{equation}
\operatorname{rank}(Q)\le mr_{\max},
\qquad
m\ge\left\lceil\frac{d}{r_{\max}}\right\rceil
\label{eq:supp-lower-bound}
\end{equation}
is necessary. For unrestricted $d\times c$ moment blocks, $r_{\max}=c$, and the bound is achievable in the arbitrary chosen-summary model. Let $m=\lceil d/c\rceil$, choose a scale $\tau>0$, form
\begin{equation}
Q=[\tau I_d,0]\in\mathbb{R}^{d\times mc},
\label{eq:supp-designed-q}
\end{equation}
and partition its columns into $m$ blocks. Then $QQ^\top=\tau^2I_d$, so $\operatorname{rank}(Q)=d$, $\sigma_{\min}(Q)=\tau$, and $\kappa_2(Q)=1$. Thus the exact noiseless query complexity is $\lceil d/c\rceil$ logical full-head responses per unknown state in this algebraic model; the statement does not exploit further dataset-realizability constraints.

For $d=512$ and $c=10$, each unknown state requires exactly 52 logical probe responses, excluding its baseline. A first attack that identifies unknown states on both sides of a deletion uses 104 probe responses, one initial baseline, two cancellations, one deletion response, and one replay response, for 109 server responses in total. If the identified pre-deletion state is cached or restored, a subsequent event may require only 52 new post-state probes plus deletion, cancellation, and replay responses, for 55 in total. A logical response here follows a complete rank-up-to-ten block; if only one rank-one constituent may precede each broadcast, the corresponding lower bound is $d=512$ responses.

\paragraph{Cumulative implementation.}
The theorem uses $Q_j$ as the total perturbation present when $W_j$ is observed. Given chosen target totals with $Q_0=0$, a cumulative implementation submits $D_j=Q_j-Q_{j-1}$ and, after observing the final response, submits only $-Q_m$.
\section{Finite-Precision Recovery}
\paragraph{Proposition 1 (direct response-error bounds).}
Suppose the observed response stack is
\begin{equation}
\widetilde R=R+E=HQ+E,
\label{eq:supp-noisy-r}
\end{equation}
where $Q$ is known exactly and has full row rank. If $\widetilde W_j=W_j+N_j$ and $\widetilde W_0=W_0+N_0$, then the $j$th block of $E$ is $N_j-N_0$; hence $E$ includes error in both probe responses and the shared baseline. Then
$\widehat H=\widetilde RQ^\dagger$ satisfies
\begin{equation}
\widehat H-H=EQ^\dagger,
\qquad
\|\widehat H-H\|_F
\le\frac{\|E\|_F}{\sigma_{\min}(Q)}.
\label{eq:supp-noise-bound}
\end{equation}
If $\widetilde R$ has full row rank, the direct estimator
\begin{equation}
\widehat A=Q\widetilde R^\dagger
\label{eq:supp-direct-a}
\end{equation}
satisfies
\begin{equation}
\widehat A-A=-AE\widetilde R^\dagger,\qquad
\|\widehat A-A\|_F
\le\frac{\|A\|_2\|E\|_F}{\sigma_{\min}(\widetilde R)}.
\label{eq:supp-a-bound}
\end{equation}

\paragraph{Proof.}
Since $QQ^\dagger=I_d$, $\widehat H-H=EQ^\dagger$, and the first bound follows from $\|Q^\dagger\|_2=1/\sigma_{\min}(Q)$. Also $Q=AR=A(\widetilde R-E)$. Using $\widetilde R\widetilde R^\dagger=I_d$ gives
\[
\widehat A-A=A(\widetilde R-E)\widetilde R^\dagger-A
=-AE\widetilde R^\dagger,
\]
which proves the direct-$A$ bound.

Let $\widehat A_{\mathrm{sym}}=\operatorname{sym}(\widehat A_{\mathrm{raw}})$. Since symmetrization is the orthogonal projection onto the symmetric matrices and $A$ is symmetric,
\begin{equation}
\|\widehat A_{\mathrm{sym}}-A\|_F
\le \|\widehat A_{\mathrm{raw}}-A\|_F.
\label{eq:supp-sym-projection}
\end{equation}
We report the equation residual before and after this projection, together with the minimum eigenvalue of $\widehat A_{\mathrm{sym}}$. Any positive-definite projection is disclosed separately. Although the probe scale $\tau$ does not change the condition number of the designed $Q$, it controls signal-to-roundoff and cancellation drift.

\section{Algebraically Admissible Moment Probes}
\paragraph{Lemma 1 (arbitrary algebraic moment probes).}
Let $e_k$ be the one-hot vector for class $k$. For any desired $Q=[q_1,\ldots,q_c]\in\mathbb R^{d\times c}$, choose $u_k=q_k/2$. An add summary induced algebraically by $(u_k,e_k)$ and a delete summary induced by $(-u_k,e_k)$ satisfy
\begin{equation}
u_ku_k^\top-(-u_k)(-u_k)^\top=0,
\end{equation}
\begin{equation}
u_ke_k^\top-(-u_k)e_k^\top=q_ke_k^\top.
\end{equation}
Summing over $k=1,\ldots,c$ realizes $Q$ with zero net Gram change. Each constituent Gram is positive semidefinite and each constituent moment is one-hot-label consistent. This proves algebraic admissibility only: it does not imply that both signed features are outputs of the shared encoder or that the deleted item belongs to the malicious client.

The designed probe in Equation~\eqref{eq:supp-designed-q} is implemented by assigning its columns to these algebraic pairs. Its scale $\tau$ does not affect the condition number, but it affects signal-to-roundoff and cancellation drift.

\paragraph{Lemma 2 (same-feature, different-label probes).}
For $u_1,\ldots,u_{c-1}\in\mathbb R^d$, add $(u_k,e_k)$ and delete $(u_k,e_c)$. The net Gram is zero and the net moment is
\begin{equation}
\smash{Q=[u_1,\ldots,u_{c-1},-\textstyle\sum_{k=1}^{c-1}u_k]},
\qquad Q\mathbf 1_c=0.
\label{eq:supp-label-swap}
\end{equation}
Conversely, every $Q$ satisfying $Q\mathbf 1_c=0$ has this representation. This construction can reuse genuine attacker features and avoids requiring $-u$ to be encoder-realizable, although deletion and label provenance remain unverified. Every resulting block has rank at most $c-1$. The corresponding lower bound is $\lceil d/(c-1)\rceil=57$ responses for $d=512,c=10$; actual full rank additionally depends on the attacker's feature span.
\section{Known Gram and Moment Probes}
Suppose probe $j$ creates known total Gram and moment perturbations $(P_j,Q_j)$ and returns
\begin{equation}
W_j=(A+P_j)^{-1}(G+Q_j),
\qquad P_j=P_j^\top,\quad A+P_j\succ0.
\end{equation}
Define $X_j=W_j-W_0$ and $Z_j=Q_j-P_jW_j$, and stack these blocks as $X$ and $Z$. Subtracting $AW_0=G$ from $(A+P_j)W_j=G+Q_j$ gives
\begin{equation}
AX=Z.
\label{eq:supp-general-probe}
\end{equation}
The state is uniquely identifiable if and only if $\operatorname{rank}(X)=d$; at full row rank, $A=ZX^\dagger$ and $G=AW_0$. For necessity, choose nonzero $v$ with $v^\top X=0$ and set $A'=A+\varepsilon vv^\top$ and $G'=A'W_0$. Then $A'X=AX=Z$, so the same observations arise from a distinct feasible ridge state. For a cumulative attacker-owned batch $(F_j,Y_j)$, $P_j=F_j^\top F_j$ and $Q_j=F_j^\top Y_j$, giving
\begin{equation}
Z_j=F_j^\top(Y_j-F_jW_j).
\end{equation}
Let $F_{\mathrm{attacker}}$ vertically stack all distinct attacker feature rows used in the sequence. Because $A$ is invertible, $\operatorname{rank}(X)=\operatorname{rank}(Z)\le\operatorname{rank}(F_{\mathrm{attacker}})$. Thus at least $d$ examples and full attacker feature rank are necessary, but not sufficient: the residuals $Y_j-F_jW_j$ and the batching sequence also affect $Z$. Label diversity is not required by this bound. The experiments below remove each cumulative addition sequence before continuing.

\paragraph{Response-rounding error for genuine additions.}
Let the returned baseline and probe heads be $\widetilde W_0=W_0+N_0$ and $\widetilde W_j=W_j+N_j$. Then $\widetilde X_j=X_j+N_j-N_0$ and $\widetilde Z_j=Z_j-P_jN_j$. If $\widetilde X$ has full row rank and $B$ stacks the blocks
\begin{equation}
B_j=-(A+P_j)N_j+AN_0,
\end{equation}
the estimator $\widehat A=\widetilde Z\widetilde X^\dagger$ obeys
\begin{equation}
\widehat A-A=B\widetilde X^\dagger,\qquad
\|\widehat A-A\|_F\le
\frac{\|B\|_F}{\sigma_{\min}(\widetilde X)}.
\label{eq:supp-genuine-rounding}
\end{equation}
Indeed, $\widetilde Z-A\widetilde X=B$ and full row rank gives $\widetilde X\widetilde X^\dagger=I_d$, which proves both the identity and the bound.
Thus cumulative Gram matrices enter the rounding-error numerator, while a better-conditioned response stack reduces the error. Increasing the probe size can affect both terms and is not unconditionally beneficial.
The data-derived implementation applies the same numerical policy as Algorithm~1: it requires full numerical rank of both $\widetilde X$ and $\widetilde Z$, symmetrizes the direct estimates $\widehat A=\widetilde Z\widetilde X^\dagger$ and $\widehat H=\widetilde X\widetilde Z^\dagger$, and requires both to be positive definite. The $H$ estimate is only an agreement diagnostic; it is not inverted to obtain $A$. Cancellation residuals are reported separately and do not determine identification success.

\section{Deleted-Update Recovery and Replay}
\paragraph{Proposition 2 (exact deletion recovery and replay).}
Let two identified states immediately bracket one deletion:
\begin{equation}
A_{\mathrm{pre}}=S_{\mathrm{pre}}+\gamma I,
\qquad
A_{\mathrm{post}}=S_{\mathrm{post}}+\gamma I.
\end{equation}
Since the same fixed regularizer appears in both states,
\begin{equation}
S_{\mathrm{del}}=A_{\mathrm{pre}}-A_{\mathrm{post}}.
\label{eq:supp-deleted-s}
\end{equation}
Moreover $G=AW$, so
\begin{equation}
G_{\mathrm{del}}
=A_{\mathrm{pre}}W_{\mathrm{pre}}
-A_{\mathrm{post}}W_{\mathrm{post}}.
\label{eq:supp-deleted-g}
\end{equation}
The deletion identities follow from
$S_{\mathrm{post}}=S_{\mathrm{pre}}-S_{\mathrm{del}}$,
$G_{\mathrm{post}}=G_{\mathrm{pre}}-G_{\mathrm{del}}$, and $G=AW$.
Adding the recovered pair to the post-deletion ledger yields
\[
A_{\mathrm{post}}+S_{\mathrm{del}}=A_{\mathrm{pre}},\qquad
G_{\mathrm{post}}+G_{\mathrm{del}}=G_{\mathrm{pre}},
\]
so the replayed head is exactly $W_{\mathrm{pre}}$. If several events occur between the states, the formulas recover only their signed aggregate net change. If $\gamma$ changes, the Gram difference is contaminated by $(\gamma_{\mathrm{pre}}-\gamma_{\mathrm{post}})I$.

If the two Gram-state estimates have errors $E_{\mathrm{pre}}$ and $E_{\mathrm{post}}$, then
\begin{equation}
\widehat S_{\mathrm{del}}-S_{\mathrm{del}}
=E_{\mathrm{pre}}-E_{\mathrm{post}},
\end{equation}
and therefore
\begin{equation}
\mathrm{RelErr}(S_{\mathrm{del}})
\le\frac{\|E_{\mathrm{pre}}\|_F+\|E_{\mathrm{post}}\|_F}
{\|S_{\mathrm{del}}\|_F}.
\label{eq:supp-deletion-amplification}
\end{equation}
For comparable state-estimation errors, smaller deletion blocks are consequently harder to recover. This explains the observed ordering from class to client to sample deletion without claiming that deletion size is the only source of error.

If the observed baselines are $\widetilde W_{\mathrm{pre}}=W_{\mathrm{pre}}+N_{\mathrm{pre}}$ and $\widetilde W_{\mathrm{post}}=W_{\mathrm{post}}+N_{\mathrm{post}}$, the deleted-moment error is exactly
\begin{equation}
\begin{aligned}
\widehat G_{\mathrm{del}}-G_{\mathrm{del}}
&=E_{\mathrm{pre}}W_{\mathrm{pre}}-E_{\mathrm{post}}W_{\mathrm{post}}\\
&\quad+\widehat A_{\mathrm{pre}}N_{\mathrm{pre}}
-\widehat A_{\mathrm{post}}N_{\mathrm{post}}.
\end{aligned}
\label{eq:supp-deleted-g-error}
\end{equation}
For a deleted sample with feature $f$, Proposition~3 therefore guarantees label recovery whenever the Frobenius norm of this right-hand side is below $\|f\|_2/2$.

\paragraph{Replay error from estimated states.}
Let $\widehat A_{\mathrm{pre}}=A_{\mathrm{pre}}+E_{\mathrm{pre}}$ and $\widehat A_{\mathrm{post}}=A_{\mathrm{post}}+E_{\mathrm{post}}$. Replaying the corresponding estimated deleted block gives
\begin{equation}
W_{\mathrm{replay}}-W_{\mathrm{pre}}
=\bigl(A_{\mathrm{pre}}+E_{\mathrm{pre}}-E_{\mathrm{post}}\bigr)^{-1}
E_{\mathrm{post}}(W_{\mathrm{pre}}-W_{\mathrm{post}}).
\label{eq:supp-replay-error}
\end{equation}
To derive the identity, the replayed state is
$\widehat A_{\mathrm{replay}}=A_{\mathrm{pre}}+E_{\mathrm{pre}}-E_{\mathrm{post}}$ and its moment is
$\widehat G_{\mathrm{replay}}=A_{\mathrm{pre}}W_{\mathrm{pre}}+E_{\mathrm{pre}}W_{\mathrm{pre}}-E_{\mathrm{post}}W_{\mathrm{post}}$.
Subtracting $\widehat A_{\mathrm{replay}}W_{\mathrm{pre}}$ from this moment gives $E_{\mathrm{post}}(W_{\mathrm{pre}}-W_{\mathrm{post}})$; left-multiplication by $\widehat A_{\mathrm{replay}}^{-1}$ yields the result.
Thus the pre-state error enters only through the inverse, while the post-state error also appears in the numerator. With rounded baselines $\widetilde W_{\mathrm{pre}}=W_{\mathrm{pre}}+N_{\mathrm{pre}}$ and $\widetilde W_{\mathrm{post}}=W_{\mathrm{post}}+N_{\mathrm{post}}$, the numerator additionally contains $\widehat A_{\mathrm{pre}}N_{\mathrm{pre}}-\widehat A_{\mathrm{post}}N_{\mathrm{post}}$.

\section{Passive Heads Do Not Identify an Aggregate Deletion}
Fix observed heads $W_{\mathrm{pre}}$ and $W_{\mathrm{post}}$. For any $A_{\mathrm{post}}\succeq\gamma I$ and any $D\succeq0$, define
\begin{equation}
A_{\mathrm{pre}}=A_{\mathrm{post}}+D,\qquad
G_{\mathrm{post}}=A_{\mathrm{post}}W_{\mathrm{post}},\qquad
G_{\mathrm{pre}}=A_{\mathrm{pre}}W_{\mathrm{pre}}.
\end{equation}
These feasible ridge states produce the same two observed heads, while different choices of $D$ give different deleted Gram blocks $S_{\mathrm{del}}=D$ and moments $G_{\mathrm{del}}=G_{\mathrm{pre}}-G_{\mathrm{post}}$. Hence passive pre/post heads do not identify a general aggregate deletion over the algebraic state class. This statement concerns unrestricted aggregate blocks; additional single-sample structure can make passive reconstruction possible.

\section{Sample-Level Leakage}
\paragraph{Proposition 3 (exact and robust label/feature recovery).}
For a deleted sample with nonzero feature $f\in\mathbb R^d$ and one-hot label $e_k$,
\[
\Delta G=fe_k^\top.
\]
Thus the unique nonzero column identifies $k$ and equals $f$. Under
$\widehat{\Delta G}=fe_k^\top+E$, define
\begin{equation}
\widehat k=\arg\max_{\ell\in\{1,\ldots,c\}}
\|\widehat{\Delta G}_{:\ell}\|_2,\qquad
\widehat f=\widehat{\Delta G}_{:\widehat k}.
\label{eq:supp-sample-rule}
\end{equation}
If
\begin{equation}
\|E_{:k}\|_2+\max_{\ell\ne k}\|E_{:\ell}\|_2<\|f\|_2,
\label{eq:supp-sample-margin}
\end{equation}
then $\widehat k=k$ and $\|\widehat f-f\|_2=\|E_{:k}\|_2$. Indeed,
$\|f+E_{:k}\|_2\ge\|f\|_2-\|E_{:k}\|_2$, whereas every incorrect column has norm $\|E_{:\ell}\|_2$. The simpler condition $\|E\|_F<\|f\|_2/2$ is sufficient.

\paragraph{Lemma 3 (information in a rank-one Gram).}
If $f\ne0$ and $gg^\top=ff^\top$, then $g=\pm f$. Both outer products have the same one-dimensional column space, so $g=af$; equality then gives $a^2=1$. A single-sample Gram therefore identifies the feature up to one global sign, not an arbitrary orthogonal transformation.

\section{Why Head Restoration Is Not a State Certificate}
\paragraph{Lemma 4 (head-equivalent ridge states).}
For a state $(A,G)$ with head $W=A^{-1}G$ and any nonzero $C\succeq0$, or more generally any symmetric $C$ satisfying $S+C\succeq0$, define
\[
A'=A+C,\qquad G'=G+CW.
\]
Then $A'=(S+C)+\gamma I$ remains a feasible ridge state and $(A+C)W=G+CW$, so $(A',G')$ has the same head. Therefore
$W_{\mathrm{final}}\approx W_0$ is only a numerical consistency check; it cannot certify equality of the hidden ledgers or the absence of an intervening honest update.

\section{Cumulative-Probe State-Recovery Algorithm}
Numerical rank counts singular values larger than $\varepsilon_{\mathrm{rank}}\sigma_{\max}$.
\textsc{SubmitMomentProbe} expands its argument into the constituent add/delete summaries of the algebraic probe construction; it does not modify the server ledger directly.
\begin{algorithm}[tbp]
\caption{Direct identification of one stable ridge state from cumulative moment probes. Each response is paired with its known total perturbation. The procedure cancels once with $-Q_m$, recovers $A$ and $H$ independently by SVD/QR solves, and reports rather than hides symmetry, definiteness, equation-residual, and agreement failures.}
\label{alg:supp-cumulative}
\begin{algorithmic}[1]
\State \textbf{Input:} baseline head $W_0$, increments $\{\delta Q_j\}_{j=1}^{m}$, rank tolerance $\varepsilon_{\mathrm{rank}}$, denominator floor $\varepsilon_{\mathrm{den}}$
\State \textbf{Output:} $\widehat A,\widehat H,\widehat G$, numerical diagnostics, and identification flag
\State $Q_{\mathrm{tot}}\gets0$
\For{$j=1,\ldots,m$}
  \State $Q_{\mathrm{tot}}\gets Q_{\mathrm{tot}}+\delta Q_j$
  \State $W_j\gets\textsc{SubmitMomentProbe}(\delta Q_j)$
  \State $Q_j\gets Q_{\mathrm{tot}}$; $R_j\gets W_j-W_0$
\EndFor
\State $Q\gets[Q_1,\ldots,Q_m]$; $R\gets[R_1,\ldots,R_m]$
\State $Q_m\gets Q_{\mathrm{tot}}$; $W_{\mathrm{final}}\gets\textsc{SubmitMomentProbe}(-Q_m)$
\If{$\operatorname{rank}_{\varepsilon_{\mathrm{rank}}}(Q)<d$ or $\operatorname{rank}_{\varepsilon_{\mathrm{rank}}}(R)<d$}
  \State \Return failure: state is not identifiable
\EndIf
\State $\widehat A_{\mathrm{raw}}\gets QR^\dagger$; $\widehat H_{\mathrm{raw}}\gets RQ^\dagger$ using SVD/QR
\State $e_{AR}\gets\|\widehat A_{\mathrm{raw}}R-Q\|_F/\max(\|Q\|_F,\varepsilon_{\mathrm{den}})$
\State $e_{HQ}\gets\|R-\widehat H_{\mathrm{raw}}Q\|_F/\max(\|R\|_F,\varepsilon_{\mathrm{den}})$
\State Record relative asymmetry of $\widehat A_{\mathrm{raw}}$ and $\widehat H_{\mathrm{raw}}$
\State $\widehat A\gets(\widehat A_{\mathrm{raw}}+\widehat A_{\mathrm{raw}}^\top)/2$
\State $\widehat H\gets(\widehat H_{\mathrm{raw}}+\widehat H_{\mathrm{raw}}^\top)/2$
\State $\lambda_A\gets\lambda_{\min}(\widehat A)$; $\lambda_H\gets\lambda_{\min}(\widehat H)$
\If{$\lambda_A\le0$ or $\lambda_H\le0$}
  \State \Return failure: unstable recovered state
\EndIf
\State $e_{AH}\gets\|\widehat A\widehat H-I_d\|_F/\sqrt d$
\State $\widehat G\gets\widehat A W_0$
\State $e_{\mathrm{rec}}\gets
\|W_{\mathrm{final}}-W_0\|_F/
\max(\|W_0\|_F,\varepsilon_{\mathrm{den}})$
\State \Return estimates, all diagnostics, and identification success
\end{algorithmic}
\end{algorithm}
The implementation sets $\varepsilon_{\mathrm{rank}}=10^{-10}$ and $\varepsilon_{\mathrm{den}}=10^{-15}$. It symmetrizes both raw direct estimates but does not project either onto the positive-definite cone: a nonpositive minimum eigenvalue is a failed identification. The cancellation-head residual $e_{\mathrm{rec}}$ is diagnostic and does not determine identification success. In simulation, where hidden ledgers are available to the evaluator but not the attacker, cancellation is additionally checked using
\begin{equation}
e_S=\frac{\|S_{\mathrm{final}}-S_0\|_F}{\max(\|S_0\|_F,\varepsilon_{\mathrm{den}})},\quad
e_G=\frac{\|G_{\mathrm{final}}-G_0\|_F}{\max(\|G_0\|_F,\varepsilon_{\mathrm{den}})},
\label{eq:supp-ledger-residuals}
\end{equation}
together with $e_W=\|W_{\mathrm{final}}-W_0\|_F/\max(\|W_0\|_F,\varepsilon_{\mathrm{den}})$. Equality of heads alone does not certify equality of ledgers.

The true deleted Gram block is positive semidefinite, but numerical recovery need not preserve this property. The evaluation therefore reports $\lambda_{\min}(\widehat S_{\mathrm{del}})$ and compares replay with the raw block against replay with its eigenvalue-clipped projection $\Pi_{\mathrm{PSD}}(\widehat S_{\mathrm{del}})$. The raw replay is primary; projected replay is a diagnostic for a server that performs a basic PSD check.

\section{Experimental Protocol and Additional Results}
The experiments were conducted on workstation machines equipped with multi-core CPUs and GPUs. The first machine used an AMD Ryzen 9 5900X GPU (12 cores, 24 threads) with 62 GB of system memory and an NVIDIA RTX A6000 GPU with 48 GB of VRAM. The second machine was configured with an Intel Core i9-11900K CPU (8 cores, 16 threads) and an NVIDIA RTX 3090 GPU with 24 GB of VRAM. The third machine used an Intel Core i7-14700K CPU (20 cores, 28 threads), 62 GB of system memory, and an NVIDIA RTX A5000 GPU with 24 GB of VRAM. For the most compute-intensive experiments, including DINOv2 feature extraction and numerical-rank checks, we additionally employed a multi-GPU server with dual AMD EPYC 7742 CPUs (128 cores, 256 threads in total), 1 TB of system memory, and eight NVIDIA A100 SXM4 GPUs, each with 80 GB of VRAM. 

Images are resized to $224\times224$ pixels and normalized with the ImageNet-1K mean and standard deviation; MNIST images are first converted to three channels. The primary encoder is ImageNet-1K ResNet-18 with its final fully connected layer replaced by the identity, so the feature is the 512-dimensional output after global average pooling. Feature extraction uses batches of 256. The ridge head has no intercept. The designed sample experiment selects 100 targets without replacement using seed 7; the class experiment deletes each of the ten classes; and the client experiment deletes all 50 clients in each of 20 partitions seeded 0--19. For each of five data-derived seeds, sample targets are 100 nonattacker examples selected without replacement, class targets are all ten classes, and client targets are 20 nonattacker clients. The data-derived and DINOv2 checks use seeds 0--4.

The server stores its ledger and solves the ridge system in float64. ``Float32 heads'' means that only returned heads and their baselines are rounded to float32 before the attacker receives them; table abbreviations f64 and f32 denote these two response precisions. Numerical rank initially uses relative tolerance $10^{-10}$; a tolerance sensitivity check is reported below. Each dataset is divided among 50 clients with Dirichlet concentration $\alpha=0.05$; the assignment routine makes at most 1,000 attempts and stops with an error rather than accepting a split with fewer than ten samples for any client. Every evaluated split passes this check.

For each data-derived split, client 0 is used if its features have rank 512; otherwise the lowest-index full-rank client is selected. The selected MNIST clients contain 867--4,309 examples and the selected CIFAR-10 clients 779--5,342; 25--33 and 25--28 of the 50 clients, respectively, satisfy the full-feature-rank gate. All selected examples are permuted and assigned without overlap to 104 batches, giving 8--42 samples per MNIST batch and 7--52 per CIFAR-10 batch. The same examples are reused for the post-deletion sequence. They are also present in the initial training ledger, so the additions are attacker-owned and encoder-realizable but are duplicate submissions rather than held-out new records. The interface does not reject duplicates. A first complete attack uses 208 addition messages, two cancellations, one honest deletion, and one replay: 212 client messages and 213 server responses including the initial baseline.

For each deletion, the program identifies and cancels a pre-deletion probe sequence, applies the deletion through the server interface, identifies and cancels a post-deletion sequence, and replays the recovered block from the actual post-cancellation state. The attacker uses only returned heads and its known probe summaries. The simulator's true ledger is used to perform the designated deletion and to score recovery, cancellation, and replay. Errors are averaged only over successful identifications; failures remain in the denominator of every success count.

\subsection{Designed-Probe Precision}
Table~\ref{tab:supp-designed-precision} gives the float32 counterpart of the main float64 table. State identification passes in every trial, but differencing two estimated states amplifies error relative to a small deletion. Designed float32 probes therefore retain all CIFAR-10 labels and 92 of 100 MNIST labels, while aggregate errors increase with finer deletion size.

\begin{table}[tbp]
\centering
\caption{Recovery when the server returns float32 heads for the deterministic designed probe with 52 responses per state. Relative errors are mean $\pm$ standard deviation over successful attacks. The descriptive 95\% Clopper--Pearson intervals are $[0.8484,0.9648]$ for MNIST and $[0.9638,1]$ for CIFAR-10; they treat trials sharing one server state as independent Bernoulli outcomes.}
\label{tab:supp-designed-precision}
\footnotesize
\setlength{\tabcolsep}{5pt}
\begin{tabular}{@{}llcccc@{}}
\toprule
Data & Deletion & Success & Labels & $\mathrm{RelErr}(\Delta G)$ & $\mathrm{RelErr}(\Delta S)$ \\
\midrule
MNIST & sample & $100/100$ & $92/100$ & $(1.09\!\pm\!.81)$ & $(3.14\!\pm\!2.74)$ \\
 & class & $10/10$ & -- & $(2.14\!\pm\!1.34)\times10^{-4}$ & $(5.76\!\pm\!4.03)\times10^{-4}$ \\
 & client & $1000/1000$ & -- & $(1.17\!\pm\!3.01)\times10^{-2}$ & $(2.89\!\pm\!7.67)\times10^{-2}$ \\
\addlinespace
CIFAR-10 & sample & $100/100$ & $100/100$ & $(2.48\!\pm\!1.82)\times10^{-1}$ & $(6.39\!\pm\!5.22)\times10^{-1}$ \\
 & class & $10/10$ & -- & $(5.68\!\pm\!2.90)\times10^{-5}$ & $(1.62\!\pm\!.89)\times10^{-4}$ \\
 & client & $1000/1000$ & -- & $(2.85\!\pm\!8.52)\times10^{-3}$ & $(6.74\!\pm\!18.27)\times10^{-3}$ \\
\bottomrule
\end{tabular}
\end{table}

Although the designed perturbation stack has condition number one at $\tau=10^4$, the returned response stack does not. Its condition number is $1.064\times10^6$ on MNIST and $4.148\times10^4$ on CIFAR-10. In float64, the corresponding direct state errors are $1.506\times10^{-12}$ and $6.549\times10^{-13}$. The response stack, rather than the designed input alone, therefore controls finite-precision recovery.

\subsection{Attacker-Data Additions}
Table~\ref{tab:supp-genuine} reports the complete data-derived deletion summary. Each row pools five probe seeds. The $m=52$ experiment targets class deletion because removing a class is the sharpest test of whether the post-deletion response stack keeps full rank. At $m=104$ and tolerance $10^{-10}$, all sample, class, and client stacks attain rank 512. The float32 results show that a successful numerical-rank test does not guarantee accurate identification.

\begin{table}[tbp]
\centering
\caption{Recovery with cumulative attacker-data additions. Errors are mean $\pm$ standard deviation over successful attacks only; failed rank tests are excluded from these means but included in ``Success.'' The displayed Clopper--Pearson intervals treat pooled trials as independent and are descriptive because trials share probe states within each of five seeds.}
\label{tab:supp-genuine}
\footnotesize
\setlength{\tabcolsep}{4pt}
\begin{tabular}{@{}lllcccc@{}}
\toprule
Data & Heads, $m$ & Deletion & Success & Labels & $\mathrm{RelErr}(\Delta G)$ & $\mathrm{RelErr}(\Delta S)$ \\
\midrule
MNIST & f64, 52 & class & $14/50$ & -- & $(4.57\!\pm\!4.11)\times 10^{-9}$ & $(5.95\!\pm\!4.79)\times 10^{-9}$ \\
CIFAR-10 & f64, 52 & class & $16/50$ & -- & $(3.65\!\pm\!2.86)\times 10^{-10}$ & $(6.28\!\pm\!5.46)\times 10^{-10}$ \\
\addlinespace
MNIST & f64, 104 & sample & $500/500$ & $500/500$ $[.9926,1]$ & $(1.55\!\pm\!1.78)\times 10^{-6}$ & $(1.62\!\pm\!2.02)\times 10^{-6}$ \\
 & & class & $50/50$ & -- & $(2.87\!\pm\!3.30)\times 10^{-10}$ & $(3.73\!\pm\!4.78)\times 10^{-10}$ \\
 & & client & $100/100$ & -- & $(1.08\!\pm\!3.02)\times 10^{-8}$ & $(8.76\!\pm\!19.49)\times 10^{-9}$ \\
CIFAR-10 & f64, 104 & sample & $500/500$ & $500/500$ $[.9926,1]$ & $(9.43\!\pm\!6.36)\times 10^{-8}$ & $(1.24\!\pm\!.89)\times 10^{-7}$ \\
 & & class & $50/50$ & -- & $(2.09\!\pm\!1.35)\times 10^{-11}$ & $(2.97\!\pm\!1.91)\times 10^{-11}$ \\
 & & client & $100/100$ & -- & $(1.36\!\pm\!3.16)\times 10^{-9}$ & $(1.85\!\pm\!4.25)\times 10^{-9}$ \\
\addlinespace
MNIST & f32, 104 & sample & $500/500$ & $49/500$ $[.0734,.1275]$ & $68.4\!\pm\!97.7$ & $143\!\pm\!254$ \\
 & & class & $50/50$ & -- & $(1.32\!\pm\!1.86)\times 10^{-2}$ & $(2.74\!\pm\!4.70)\times 10^{-2}$ \\
 & & client & $100/100$ & -- & $.393\!\pm\!1.03$ & $.661\!\pm\!1.97$ \\
CIFAR-10 & f32, 104 & sample & $500/500$ & $183/500$ $[.3237,.4099]$ & $5.27\!\pm\!4.28$ & $10.1\!\pm\!11.1$ \\
 & & class & $50/50$ & -- & $(1.16\!\pm\!.87)\times 10^{-3}$ & $(2.42\!\pm\!2.18)\times 10^{-3}$ \\
 & & client & $100/100$ & -- & $.0842\!\pm\!.215$ & $.161\!\pm\!.528$ \\
\bottomrule
\end{tabular}
\end{table}

At $m=52$, all five pre-deletion states have full rank, but most post-class-deletion states have rank 468; the observed range is 468--512. At $m=104$, median response-stack condition numbers are $4.06\times10^5$ on MNIST and $1.43\times10^5$ on CIFAR-10. Float32 sample-label rates by seed are 7--14\% on MNIST and 12--100\% on CIFAR-10, so the pooled CIFAR-10 rate hides substantial state dependence. Full attacker feature rank remains necessary but is not sufficient.

Table~\ref{tab:supp-rank-sensitivity} tests whether float32 quantization merely creates tiny singular values that pass the original $10^{-10}$ numerical-rank threshold. For each dataset and batch count, it recomputes one pre-deletion and one post-class-deletion response stack for each of five seeds. At $m=104$, all stacks remain full rank from $10^{-7}$ through $10^{-10}$. At $10^{-6}$, all CIFAR-10 stacks but only six of ten MNIST stacks are full rank. Thus the $m=104$ conclusion is stable at a tolerance comparable to float32 unit roundoff, but some MNIST directions are marginal under a ten-times-larger threshold. In all float32 tables, ``Success'' means that the estimator returned after the stated numerical-rank and positive-definiteness checks; it does not imply accurate state recovery.

\begin{table}[tbp]
\centering
\caption{Numerical-rank sensitivity of float32 attacker-data response stacks. Each row contains ten matrices: one pre-deletion and one post-class-deletion stack for each of five seeds. Entries are the number with full rank 512 under the relative singular-value threshold shown. The smallest-singular-value range includes all ten matrices.}
\label{tab:supp-rank-sensitivity}
\footnotesize
\setlength{\tabcolsep}{4pt}
\begin{tabular}{@{}llccccccl@{}}
\toprule
Data & $m$ & $10^{-6}$ & $10^{-7}$ & $10^{-8}$ & $10^{-9}$ & $10^{-10}$ & $\sigma_{\min}$ range \\
\midrule
MNIST & 52 & 0 & 2 & 6 & 7 & 7 & $2.35\times10^{-21}$--$5.29\times10^{-8}$ \\
MNIST & 104 & 6 & 10 & 10 & 10 & 10 & $1.24\times10^{-7}$--$2.38\times10^{-6}$ \\
CIFAR-10 & 52 & 1 & 5 & 6 & 6 & 6 & $2.37\times10^{-20}$--$7.23\times10^{-7}$ \\
CIFAR-10 & 104 & 10 & 10 & 10 & 10 & 10 & $6.73\times10^{-7}$--$1.08\times10^{-5}$ \\
\bottomrule
\end{tabular}
\end{table}

\subsection{Numerical Diagnostics}
Table~\ref{tab:supp-state-diagnostics} reports the state-equation checks promised in the main paper. The designed rows evaluate the common baseline state used by the deletion experiments; the attacker-data rows summarize the five float64 baseline identifications. For the latter, residuals are maxima and eigenvalues are minima, giving the least favorable value across seeds.

\begin{table}[tbp]
\centering
\caption{Float64 state-identification diagnostics. $e_{AR}$ and $e_{HQ}$ are relative residuals of the two solved matrix equations; $e_{AH}=\|\widehat A\widehat H-I\|_F/\sqrt d$ measures agreement between independently estimated regularized state and inverse. Positive minimum eigenvalues confirm both estimates pass the definiteness checks.}
\label{tab:supp-state-diagnostics}
\footnotesize
\setlength{\tabcolsep}{3.5pt}
\begin{tabular}{@{}llccccc@{}}
\toprule
Probe & Data & $e_{AR}$ & $e_{HQ}$ & $e_{AH}$ & $\lambda_{\min}(\widehat A)$ & $\lambda_{\min}(\widehat H)$ \\
\midrule
Designed & MNIST & $3.30\times10^{-12}$ & $2.07\times10^{-16}$ & $2.25\times10^{-12}$ & 23.2 & $4.05\times10^{-8}$ \\
Designed & CIFAR-10 & $4.33\times10^{-13}$ & $5.25\times10^{-16}$ & $3.42\times10^{-13}$ & 638 & $3.78\times10^{-8}$ \\
Attacker data & MNIST & $8.84\times10^{-13}$ & $1.37\times10^{-11}$ & $5.53\times10^{-11}$ & 23.2 & $4.05\times10^{-8}$ \\
Attacker data & CIFAR-10 & $9.20\times10^{-14}$ & $2.82\times10^{-12}$ & $6.53\times10^{-12}$ & 638 & $3.78\times10^{-8}$ \\
\bottomrule
\end{tabular}
\end{table}

The simulator-only cancellation checks are similarly small. Across all successful float64 attacker-data sample, class, and client experiments, the maxima of $(e_S,e_G,e_W)$ are $(5.06\times10^{-16},3.24\times10^{-16},7.89\times10^{-13})$ on MNIST and $(5.20\times10^{-16},3.61\times10^{-16},2.07\times10^{-13})$ on CIFAR-10. These values use the hidden ledgers only for evaluation; the recovery procedure never reads them.

\begin{table}[tbp]
\centering
\caption{Float64 replay and positive-semidefiniteness diagnostics for attacker-data additions. Values are means over successful trials except the minimum eigenvalue, which is the worst case. ``Raw/PSD'' compares the recovered deleted Gram block with the same block after clipping negative eigenvalues to zero. Moment error is unchanged by this projection. Replay error compares each resulting head with the pre-deletion head.}
\label{tab:supp-psd-replay}
\footnotesize
\setlength{\tabcolsep}{3.5pt}
\begin{tabular}{@{}lllccc@{}}
\toprule
Data & Level & $\min\lambda(\widehat S_{\rm del})$ & $\mathrm{RelErr}(S)$ raw/PSD & $\mathrm{RelErr}(G)$ & Replay raw/PSD \\
\midrule
MNIST & sample & $-1.42\times10^{-3}$ & $(3.22/3.10)\times10^{-11}$ & $8.73\times10^{-11}$ & $1.03\times10^{-12}/5.61\times10^{-9}$ \\
 & class & $1.06\times10^{-1}$ & $(3.95/3.95)\times10^{-11}$ & $9.05\times10^{-11}$ & $(2.36/2.36)\times10^{-10}$ \\
 & client & $-1.17\times10^{-4}$ & $(3.37/3.37)\times10^{-11}$ & $8.25\times10^{-11}$ & $3.06\times10^{-11}/7.20\times10^{-10}$ \\
CIFAR-10 & sample & $-1.06\times10^{-4}$ & $(3.31/2.99)\times10^{-12}$ & $6.73\times10^{-12}$ & $2.32\times10^{-13}/3.68\times10^{-10}$ \\
 & class & $1.50\times10^{1}$ & $(3.21/3.21)\times10^{-12}$ & $6.59\times10^{-12}$ & $(3.42/3.42)\times10^{-11}$ \\
 & client & $-4.00\times10^{-5}$ & $(3.32/3.30)\times10^{-12}$ & $6.51\times10^{-12}$ & $4.16\times10^{-12}/7.86\times10^{-11}$ \\
\bottomrule
\end{tabular}
\end{table}

Small negative eigenvalues occur for some sample and client estimates, so a server that rejects every non-PSD submitted block would reject those raw replays. Projection removes this numerical violation and leaves the recovered block accurate, but it no longer reproduces the head as closely. The projected replay errors in Table~\ref{tab:supp-psd-replay} remain below $6\times10^{-9}$ in float64.

\begin{table}[tbp]
\centering
\caption{Relative error of the recovered deleted feature, conditioned on the maximum-column rule selecting the correct sample label. Rows give the number of correctly labelled samples and the mean, median, and maximum feature error.}
\label{tab:supp-feature-error}
\footnotesize
\setlength{\tabcolsep}{5pt}
\begin{tabular}{@{}lllcccc@{}}
\toprule
Data & Heads & Correct & Mean & Median & Maximum \\
\midrule
MNIST & float64 & 500 & $4.32\times10^{-7}$ & $2.00\times10^{-7}$ & $3.95\times10^{-6}$ \\
CIFAR-10 & float64 & 500 & $2.75\times10^{-8}$ & $2.36\times10^{-8}$ & $1.99\times10^{-7}$ \\
MNIST & float32 & 49 & $4.01\times10^{1}$ & $1.58\times10^{1}$ & $2.32\times10^{2}$ \\
CIFAR-10 & float32 & 183 & $1.14$ & $2.14\times10^{-1}$ & $8.15$ \\
\bottomrule
\end{tabular}
\end{table}

\subsection{DINOv2 Encoder Robustness}
We replace ResNet-18 with the frozen DINOv2 ViT-B/14 encoder \cite{oquab2024dinov2}, increasing $d$ from 512 to 768 and the dimensional lower bound from 52 to 77 responses. This experiment identifies one state and does not rerun the deletion or decoder evaluations. Table~\ref{tab:supp-dinov2} shows the same distinction as the main experiment: the dimensional count is necessary but does not guarantee observed rank for attacker-data additions. All 104-response estimates are positive definite; their relative $G$ errors range from $1.87\times10^{-11}$ to $5.67\times10^{-11}$ on MNIST and from $3.26\times10^{-12}$ to $8.61\times10^{-12}$ on CIFAR-10.

\begin{table}[tbp]
\centering
\caption{One-state identification from attacker-data additions with frozen 768-dimensional DINOv2 ViT-B/14 features. Each row summarizes five independently seeded client splits and probe sequences. A usable attacker has 768-dimensional feature rank and at least 77 samples. Ranges are minima and maxima over the five runs; errors are omitted when numerical rank fails.}
\label{tab:supp-dinov2}
\footnotesize
\setlength{\tabcolsep}{3.5pt}
\begin{tabular}{@{}llccccc@{}}
\toprule
Data & $m$ & Success & $\operatorname{rank}(X)$ & Usable & $\kappa(X)$ & $\mathrm{RelErr}(A)$ \\
\midrule
MNIST & 77 & $0/5$ & 762--766 & 20--26 & -- & -- \\
MNIST & 104 & $5/5$ & 768 & 20--26 & $(1.51,4.34)\times10^6$ & $(4.14\times10^{-12},2.50\times10^{-11})$ \\
CIFAR-10 & 77 & $0/5$ & 746--756 & 21--26 & -- & -- \\
CIFAR-10 & 104 & $5/5$ & 768 & 21--26 & $(3.23\times10^5,1.12\times10^6)$ & $(3.17,8.53)\times10^{-12}$ \\
\bottomrule
\end{tabular}
\end{table}

The raw replay-head errors for float64 sample/class/client deletion are $1.03\times10^{-12}$, $2.36\times10^{-10}$, and $3.06\times10^{-11}$ on MNIST, and $2.32\times10^{-13}$, $3.42\times10^{-11}$, and $4.16\times10^{-12}$ on CIFAR-10. With float32 heads, they are $1.54\times10^{-7}$, $4.16\times10^{-4}$, and $5.72\times10^{-5}$ on MNIST, and $5.40\times10^{-8}$, $1.33\times10^{-4}$, and $1.62\times10^{-5}$ on CIFAR-10. A small head error does not certify a correct replayed ledger, as proved in Lemma~4.

\subsection{Designed-Probe Scale and Detectability}
For an add/delete request, the update magnitude is the combined Frobenius norm of its four submitted summary matrices. Table~\ref{tab:supp-scale} compares the largest designed request and its largest induced head change with the 99th percentiles of honest requests. All 20 deletions succeed at every scale, but most scales are conspicuous. At $\tau=10^4$, clipping the request magnitude to the honest 99th percentile retains $20/20$ success, with mean state errors $7.07\times10^{-11}$ before and $4.37\times10^{-11}$ after deletion, deleted-$S/G$ errors $3.68\times10^{-6}/1.76\times10^{-6}$, and a head change still 149 times the honest 99th percentile.

\begin{table}[tbp]
\centering
\caption{Fixed MNIST float64 scale sweep for designed probes, with 20 sample deletions per scale. $\mathrm{RelErr}(A)$ is the pre-deletion state-identification error. ``Update ratio'' and ``head ratio'' divide the largest malicious request and head change by the corresponding honest 99th percentile.}
\label{tab:supp-scale}
\footnotesize
\setlength{\tabcolsep}{4pt}
\begin{tabular}{@{}rcccccc@{}}
\toprule
$\tau$ & Success & $\mathrm{RelErr}(A)$ & $\mathrm{RelErr}(\Delta S)$ & $\mathrm{RelErr}(\Delta G)$ & Update ratio & Head ratio \\
\midrule
$10^0$  & $20/20$ & $1.57\times 10^{-9}$  & $1.00\times 10^{-4}$ & $3.46\times 10^{-5}$ & $5.15\times 10^{-3}$ & $8.15$ \\
$10^2$  & $20/20$ & $1.91\times 10^{-11}$ & $1.01\times 10^{-6}$ & $3.53\times 10^{-7}$ & $2.98\times 10^{1}$ & $8.15\times 10^{2}$ \\
$10^4$  & $20/20$ & $1.51\times 10^{-12}$ & $1.01\times 10^{-7}$ & $3.50\times 10^{-8}$ & $2.98\times 10^{5}$ & $8.15\times 10^{4}$ \\
$10^6$  & $20/20$ & $1.24\times 10^{-12}$ & $1.05\times 10^{-7}$ & $3.72\times 10^{-8}$ & $2.98\times 10^{9}$ & $8.15\times 10^{6}$ \\
$10^8$  & $20/20$ & $7.96\times 10^{-13}$ & $1.05\times 10^{-7}$ & $4.07\times 10^{-8}$ & $2.98\times 10^{13}$ & $8.15\times 10^{8}$ \\
$10^{10}$ & $20/20$ & $3.72\times 10^{-13}$ & $7.43\times 10^{-8}$ & $5.80\times 10^{-7}$ & $2.98\times 10^{17}$ & $8.15\times 10^{10}$ \\
$10^{12}$ & $20/20$ & $4.27\times 10^{-13}$ & $8.20\times 10^{-8}$ & $3.67\times 10^{-5}$ & $2.98\times 10^{21}$ & $8.15\times 10^{12}$ \\
\bottomrule
\end{tabular}
\end{table}

For attacker-data additions, batch-size-specific 99th-percentile norm thresholds are calibrated on 5,000 honest messages per dataset and evaluated on 10,000 separate honest messages. Held-out false-positive rates are 1.38\% on MNIST and 1.34\% on CIFAR-10. The thresholds flag 0 of 520 MNIST additions and 13 of 520 CIFAR-10 additions; the median probe-to-threshold ratios are 0.818 and 0.845. These per-message results exclude the two cumulative cancellations and replay and do not test temporal or frequency-based detection.

Rounding a final honest ResNet-18 ridge head from float64 to float32 changes no prediction among 10,000 test examples on either dataset. Test accuracy remains 97.15\% on MNIST and 85.94\% on CIFAR-10, and the relative head-rounding errors are $2.57\times10^{-8}$ on both datasets. This evaluates inference from the rounded broadcast, not the utility of training a server ledger in float32.

\section{Additional Reconstruction and Replay Figures}

\begin{figure}[tbp]
\centering
{\footnotesize
\setlength{\tabcolsep}{4pt}%
\begin{tabular}{m{1.7cm}m{.42\textwidth}m{.42\textwidth}}
& \multicolumn{1}{c}{MNIST} & \multicolumn{1}{c}{CIFAR-10} \\
Target & \includegraphics[width=\linewidth]{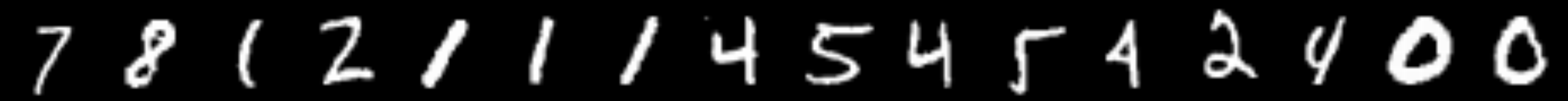}
       & \includegraphics[width=\linewidth]{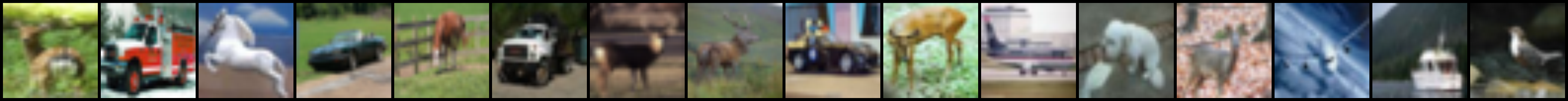} \\
$\alpha=0.05$ & \includegraphics[width=\linewidth]{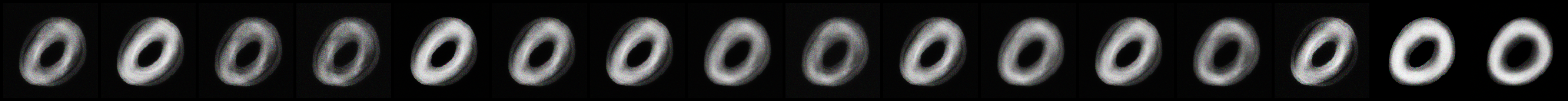}
              & \includegraphics[width=\linewidth]{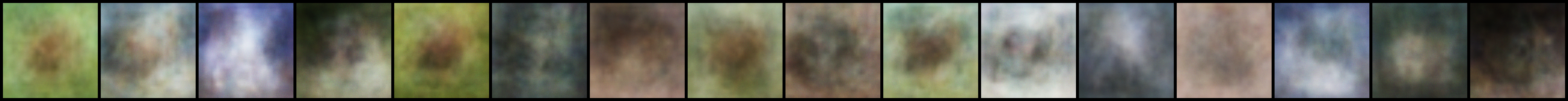} \\
$\alpha=0.5$ & \includegraphics[width=\linewidth]{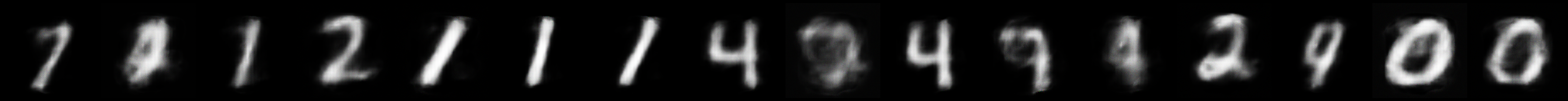}
             & \includegraphics[width=\linewidth]{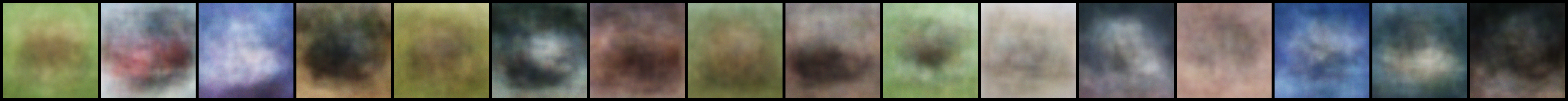} \\
IID & \includegraphics[width=\linewidth]{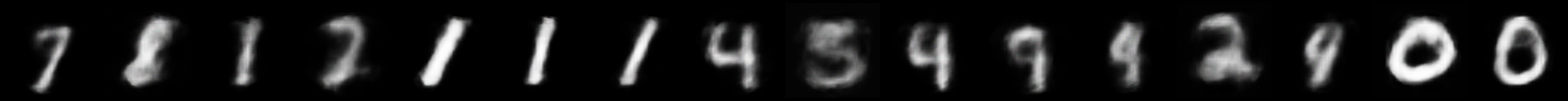}
    & \includegraphics[width=\linewidth]{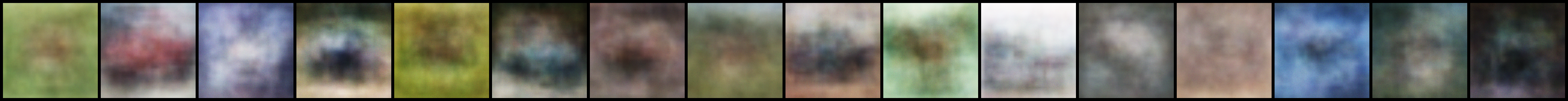} \\
Exact feature & \includegraphics[width=\linewidth]{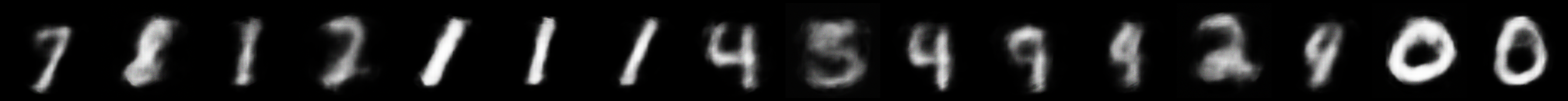}
              & \includegraphics[width=\linewidth]{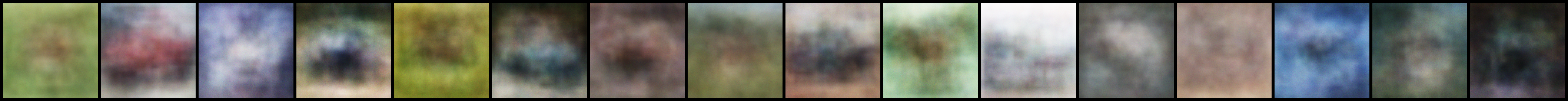} \\
\end{tabular}}
\caption{Selected qualitative decoder outputs for MNIST (middle) and CIFAR-10 (right). Rows use decoders trained only on one malicious client's local image--feature pairs under strongly non-IID ($\alpha=0.05$), moderately non-IID ($\alpha=0.5$), or independent and identically distributed (IID) client partitions. The final row decodes the exact target feature with the IID decoder and separates feature-recovery error from decoder loss. These selected examples are not a quantitative reconstruction evaluation.}
\label{fig:supp-local-decoder}
\end{figure}


\begin{figure}[tbp]
\centering
{\footnotesize
\setlength{\tabcolsep}{5pt}%
\begin{tabular}{m{1.8cm}m{.42\textwidth}m{.42\textwidth}}
& \multicolumn{1}{c}{MNIST} & \multicolumn{1}{c}{CIFAR-10} \\
Target image & \includegraphics[width=\linewidth]{mnist/mnist_row_original.png}
             & \includegraphics[width=\linewidth]{cifar/cifar10_row_original.png} \\
LDM reconstruction & \includegraphics[width=\linewidth]{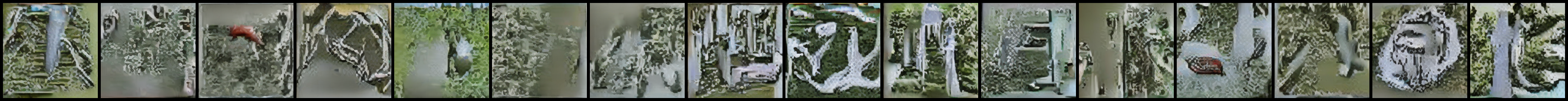}
                   & \includegraphics[width=\linewidth]{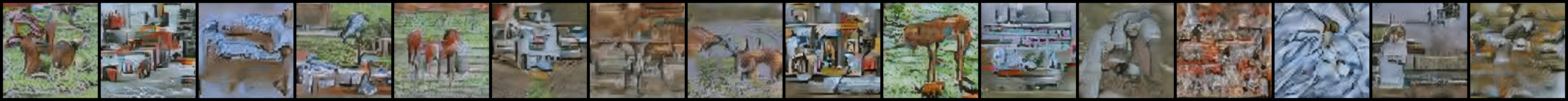} \\
\end{tabular}}
\caption{Feature-conditioned latent diffusion reconstructions for MNIST (middle column) and CIFAR-10 (right column). The top row contains target images and the bottom row contains samples generated from recovered frozen features by a model trained on auxiliary TinyImageNet data; these examples test coarse visual leakage rather than faithful pixel recovery.}
\label{fig:supp-ldm}
\end{figure}

\begin{figure}[tbp]
\centering
\includegraphics[width=.46\textwidth]{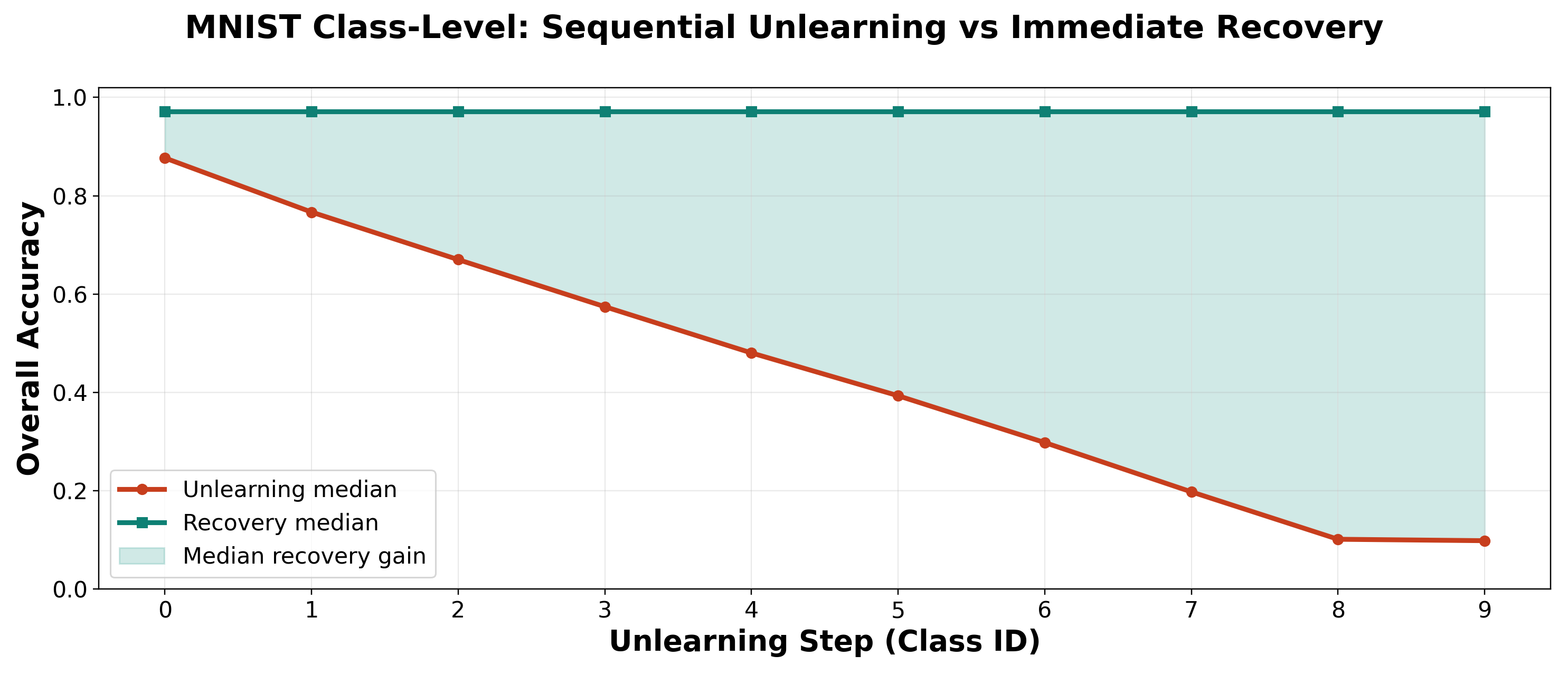}
\hfill
\includegraphics[width=.46\textwidth]{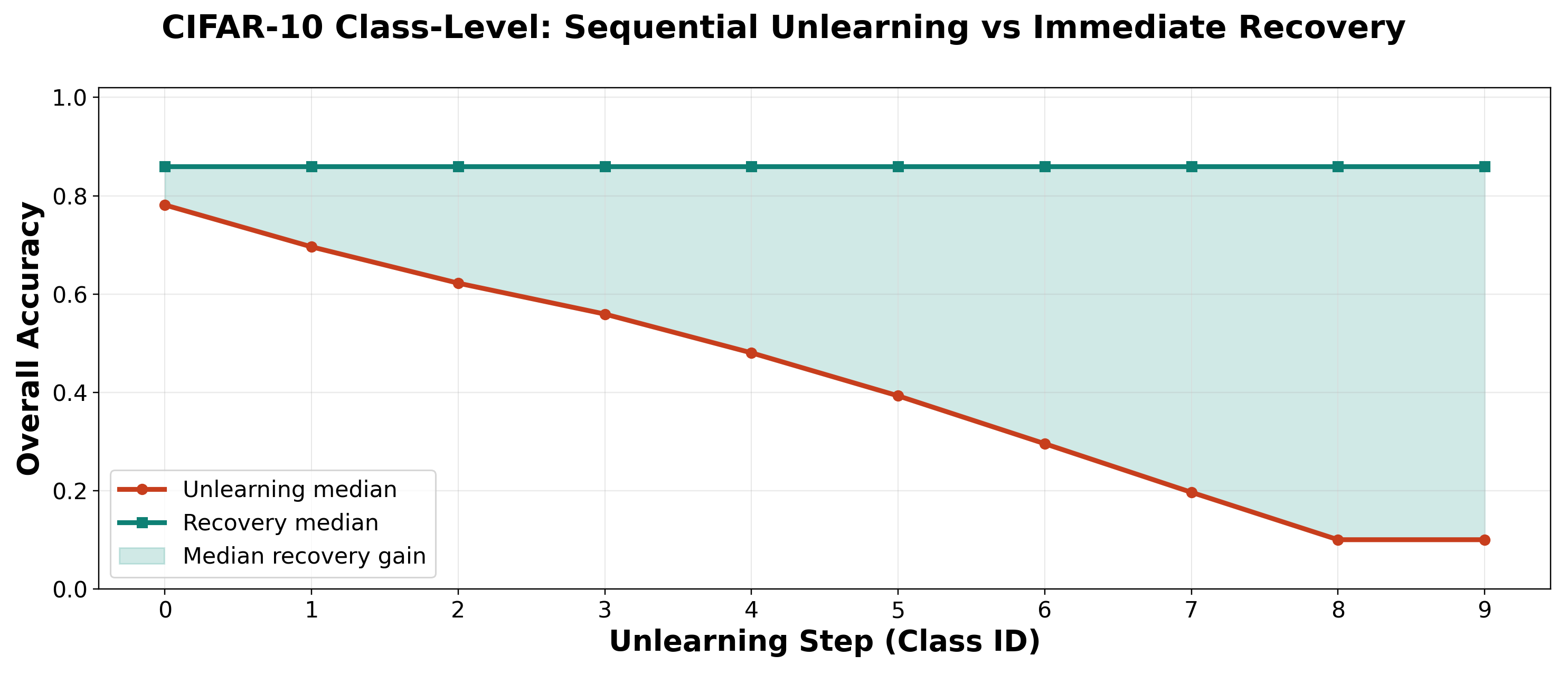}
\caption{Test accuracy during ten sequential class deletions with float64 designed probes on MNIST (left) and CIFAR-10 (right). The honest branch retains every deletion; the attacked branch immediately replays each recovered class block before the next deletion. Final honest/attacked accuracies are 9.82\%/97.15\% and 8.94\%/85.94\%, respectively.}
\label{fig:supp-class-replay}
\end{figure}

\begin{figure}[tbp]
\centering
\includegraphics[width=.46\textwidth]{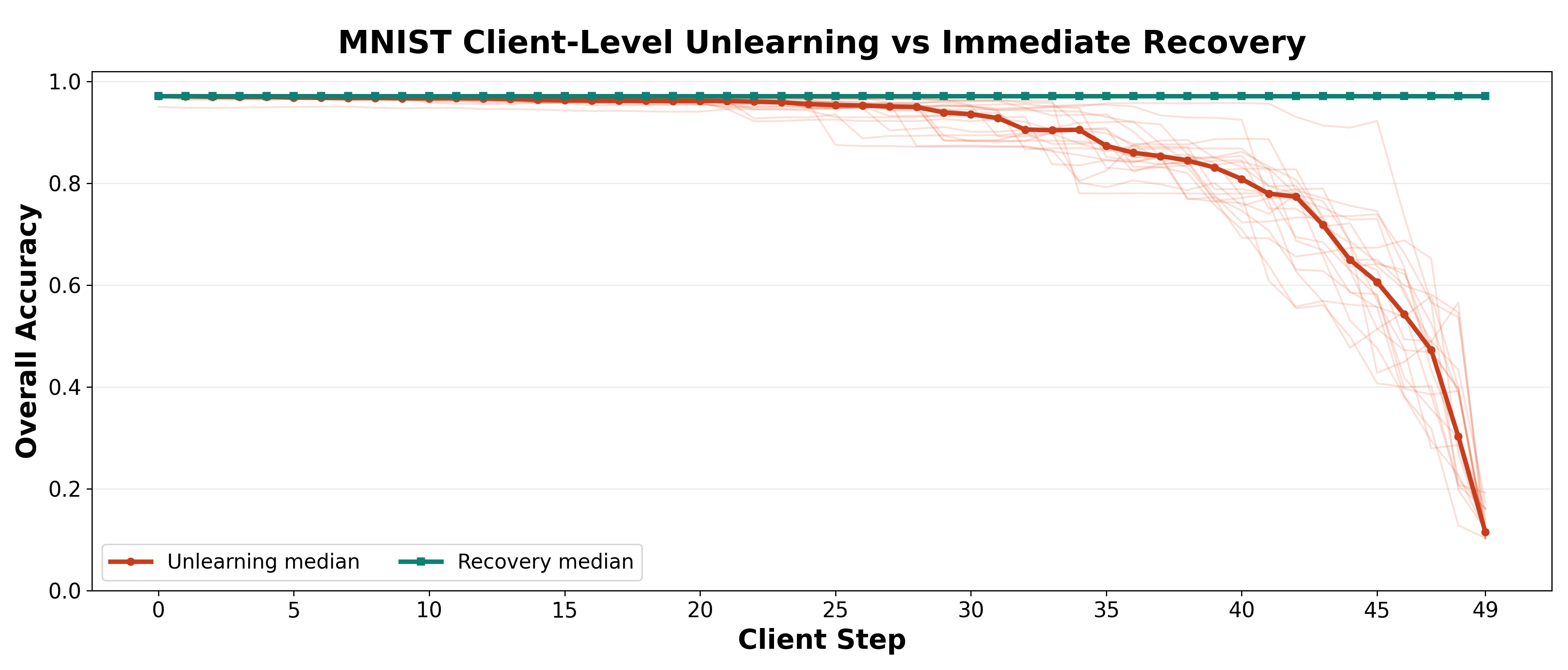}
\hfill
\includegraphics[width=.46\textwidth]{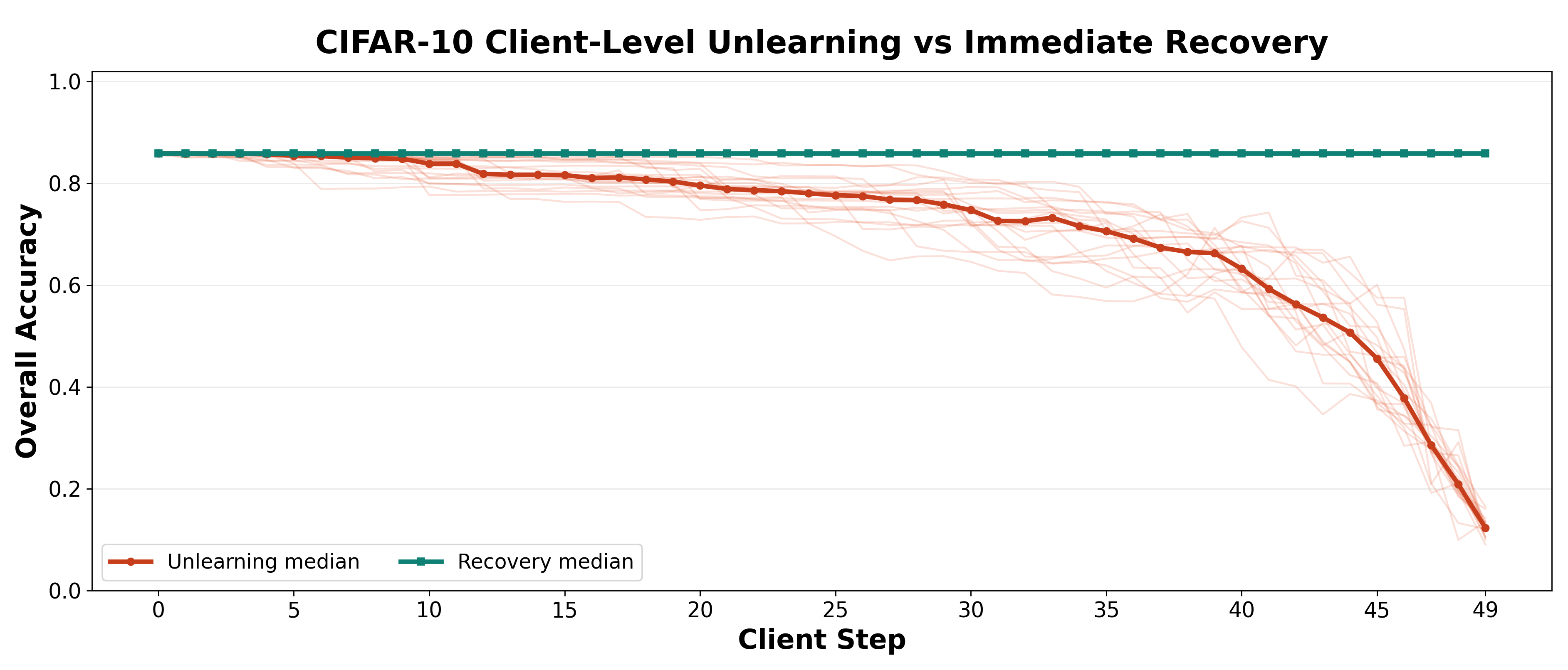}
\caption{Client-level replay over 50 sequential deletion rounds on MNIST (left) and CIFAR-10 (right). The honest branch retains every client deletion; the attacked branch immediately replays each recovered client block before the next round. Thin curves show 20 independently sampled client partitions, and emphasized curves show their medians.}
\label{fig:supp-client-replay}
\end{figure}

Recovering frozen feature representations from the head exposes an additional channel for coarse visual leakage. To illustrate this potential security risk, we train feature decoders using (1) a local decoder and (2) a decoder trained on auxiliary data. The local decoder used for each partition is trained only on the selected malicious client's actual images and their frozen features; it does not use the probe construction or any target image. Figure~\ref{fig:supp-local-decoder} shows selected outputs. We use a lightweight multilayer perceptron (MLP) for feature decoding, trained with a standard supervised pixel-level reconstruction objective. In addition to this locally trained MLP decoder, we also report reconstructions from a conditional latent diffusion decoder trained on auxiliary TinyImageNet data, which conditions a UNet-based latent denoiser using feature-derived tokens. Figure~\ref{fig:supp-ldm} shows that although the domain shift prevents high‑fidelity reconstructions, coarse visual leakage may still be visible. Overall, the reconstructions from local decoder and auxiliary data decoder demonstrate that, once a malicious client recovers frozen features, they can induce coarse visual leakage or reveal individual digit identities in MNIST examples. We include these examples only to illustrate coarse visual leakage, and do not claim faithful pixel-level recovery. 

Figure~\ref{fig:supp-class-replay} and Figure~\ref{fig:supp-client-replay} illustrate statistics replay after each class- and client-level deletion. The blue curves show test accuracy under immediate probing and reinsertion after every unlearning step, while the red curves show the effect of honest unlearning without replay. Reinserting the estimated aggregate block restores test accuracy to its pre-deletion level.
